\documentclass[a4paper,usenatbib,useAMS]{mnras}

\usepackage{newtxtext,newtxmath}

\usepackage[T1]{fontenc}

\DeclareRobustCommand{\VAN}[3]{#2}
\let\VANthebibliography\thebibliography
\def\thebibliography{\DeclareRobustCommand{\VAN}[3]{##3}\VANthebibliography}

\usepackage{graphicx}	
\usepackage{amsmath}	

\usepackage{amssymb}	
\usepackage{ulem}       

\usepackage{tikz}
\usetikzlibrary{arrows,calc,positioning,shapes.geometric}
\usepackage{comment}
\usepackage{enumitem}
\usepackage{caption}
\usepackage{subcaption}
\usepackage[T4, T1]{fontenc}
\usepackage{placeins}
\usepackage{float}
\usepackage{lineno}
\tikzstyle{All} = [rectangle, minimum width = 1cm, minimum height = 1cm,text centered,text width=2cm,draw=black]
\tikzstyle{process} = [rectangle,rounded corners, minimum width = 1cm, minimum height = 1cm,text centered,text width=1.5cm, draw=black, fill=green!30]
\tikzstyle{arrow} = [thick, ->, >=stealth]

\newcommand{\lsim}{{\;\raise0.3ex\hbox{$<$\kern-0.75em\raise-1.1ex\hbox{$\sim$}}\;}}
\newcommand{\gsim}{{\;\raise0.3ex\hbox{$>$\kern-0.75em\raise-1.1ex\hbox{$\sim$}}\;}}

\newcommand{\arf}[1]{\texttt{#1}}

\title[Dynamics of  Multiphase CGM around BGGs]{Cool and gusty, with a chance of rain: Dynamics of multiphase CGM around massive galaxies in the {\sc Romulus} simulations}

\author[Saeedzadeh et al.]{Vida Saeedzadeh$^{1}$\thanks{e-mail: \href{mailto:vidasaeedzadeh@uvic.ca}{vidasaeedzadeh@uvic.ca}},
S. Lyla Jung$^{2}$,
Douglas Rennehan$^{3}$,
Arif Babul$^{1}$\thanks{Infosys Visiting Chair Professor, Indian Institute of Science, Bangalore 560012, India.},
Michael Tremmel$^{4}$,
\newauthor Thomas R. Quinn$^{5}$,
Zhiwei Shao$^{6}$,
Prateek Sharma$^{7}$,
Lucio Mayer$^{8}$,
E. O’Sullivan$^{9}$,
S. Ilani Loubser$^{10}$
\\\\
$^{1}$Department of Physics and Astronomy, University of Victoria, 3800 Finnerty Road, Victoria, BC, V8P 1A1, Canada\\
$^{2}$Research School of Astronomy \& Astrophysics, Australian National University, Canberra, ACT 2611, Australia\\
$^{3}$Center for Computational Astrophysics, Flatiron Institute, 162 5th Ave, New York, NY 10010, USA\\
$^{4}$School of Physics, University College Cork, College Road, Cork T12 K8AF, Ireland \\
$^{5}$Astronomy Department, University of Washington, Box 351580, Seattle, WA, 98195-1580, USA\\
$^{6}$Department of Astronomy, School of Physics and Astronomy, Shanghai Jiao Tong University, Shanghai 200240, China\\
$^{7}$Department of Physics and Joint Astronomy Program, Indian Institute of Science, Bangalore 560012, India\\
$^{8}$Institute for Computational Science, University of Zürich, Winterthurerstrasse 190, 8057 Zürich, Switzerland\\
$^{9}$Center for Astrophysics | Harvard \& Smithsonian, 60 Garden Street, Cambridge, MA 02138, USA\\
$^{10}$Centre for Space Research, North-West University, Potchefstroom 2520, South Africa
}

\date{Accepted XXX. Received YYY; in original form ZZZ}

\pubyear{2023}

\begin{document}
\label{firstpage}
\pagerange{\pageref{firstpage}--\pageref{lastpage}}
\maketitle

\begin{abstract}

Using high-resolution {\sc Romulus} simulations, we explore the origin and evolution of the circumgalactic medium (CGM) in the region 0.1 $\leq \mathrm{R}/\mathrm{R}_\mathrm{500} \leq$ 1 around massive central galaxies in group-scale halos.   We find that the CGM is multiphase and highly dynamic. Investigating the dynamics, we identify seven patterns of evolution. We show that these are robust and detected consistently across various conditions. The gas cools via two pathways: (1) filamentary cooling inflows and (2) condensations forming from rapidly cooling density perturbations. In our cosmological simulations, the perturbations are mainly seeded by orbiting substructures. The condensations can form even when the median $t_\mathrm{cool} / t_\mathrm{ff}$  of the X-ray emitting gas is above 10 or 20.  Strong amplitude perturbations can provoke runaway cooling regardless of the state of the background gas. We also find perturbations whose local $t_\mathrm{cool} / t_\mathrm{ff}$ ratios drop below the threshold but which do not condense.  Rather, the ratios fall to some minimum value and then bounce. These are weak perturbations that are temporarily swept up in satellite wakes and carried to larger radii. Their $t_\mathrm{cool} / t_\mathrm{ff}$ ratios decrease because $t_\mathrm{ff}$ is increasing, not because $t_\mathrm{cool}$ is decreasing.  For structures forming hierarchically, our study highlights the challenge of using a simple threshold argument to infer the CGM's evolution.  It also highlights that the median hot gas properties are suboptimal determinants of the CGM's state and dynamics. Realistic CGM models must incorporate the impact of mergers and orbiting satellites, along with the CGM's heating and cooling cycles.

\end{abstract}





\section{Introduction}\label{Introduction}

It is largely accepted that conditions and processes arising in galaxy group environments must play a crucial role in galaxy evolution. Owing to higher galaxy density and relatively low galaxy velocity dispersion (cf. \citealt{o2017complete,werner-mernier2020review,Lovisari-Ettori2021UReview} and references therein), group galaxies, both central and satellites, have a higher vulnerability to mergers and tidal interactions.  These galaxies are also subjected to  ram pressure stripping \citep{mccarthy2008ram} and cooling flows due to their being immersed in an ocean of hot X-ray emitting plasma. After all, not only do a substantial fraction of galaxies in the Universe (>50\%; \citealt{eke2006galaxy}) reside in groups, some of the most massive galaxies in the cosmos are also forged in group environments (cf. \citealt{Rennehan2020_proto,jung2022massive} and references therein).

 In spite of this, groups have received far less attention than individual galaxies or the more massive galaxy clusters.  One reason for this is that they are difficult to detect in X-ray observations due to their low surface brightness;  they are also difficult to identify reliably in optical catalogs due to their relatively small galaxy number density contrast \citep{Pearson2017,o2017complete}.  There is change afoot, however.  In part, this is due to  recent and upcoming compilations that promise high fidelity group catalogs. These include the multiwavelength Complete Local-Volume Groups Sample (CLoGS; e.g. \citealt{o2017complete,o2018cold,kolokythas2019complete,kolokythas2022complete}) and galaxy group catalogs from the 2MASS Redshift Survey \citep{lambert20202mass}, the REFINE survey \citep{sarron2021detectifz},  
 the eROSITA Final Equatorial-Depth Survey \citep{liu2022erosita}, etc. The heightened attention is also in part due to improving group detection algorithms \citep[e.g.~][]{Ibrahem2015,Maggie2015,Yang2021,Xu2022} and  the increasingly sophisticated cosmological galaxy formation simulations that are now able to reproduce the observed stellar properties of group galaxies as well as facilitate new insights into group-scale processes and their impact on galaxies (see the recent review by \citealt{oppenheimer2021simulating} as well as the recent study by \citealt{jung2022massive} for a summary discussion of the relevant simulations.) 
 
One such set is the {\sc Romulus} suite of cosmological hydrodynamic simulations \citep{tremmel2017romulus,tremmel2019introducing,butsky2019ultraviolet,chadayammuri2020fountains,jung2022massive} comprising one uniform cosmological volume (25 Mpc per side)  and three group-scale zoom simulations.  Recently, \citet{jung2022massive} examined the kinematic and photometric properties of the brightest group galaxies (hereafter, BGGs) in {\sc Romulus} groups.
They found that the distribution of their properties,
the trends exhibited by and the correlations among them, are in very good agreement with the observations. Like the observed BGGs, the {\sc Romulus} BGGs include both quenched and star forming galaxies; early-type elliptical galaxies and late-type disk galaxies; fast-rotators and slow-rotators; etc.  \citet{jung2022massive} also follow the evolution of BGGs transforming from late- to early-type in the aftermath of mergers. More interestingly, they also find ``galaxy rejuvenation'' where quenched early-type BGGs transition into late-type star-forming galaxy (see also \citealt{
jackson2022extremely}.)

The latter is, at first glance, surprising and tempting to dismiss as an artefact, but recent results from the Multi Unit Spectroscopic Explorer (MUSE; \citealt{olivares2022gas}) suggest that the re-emergence of gaseous disks and star forming disks/rings around BGGs, at least on a small scale, is not uncommon (see also \citealt{Loubser2022,Lagos2022}).    Moreover, \citet{Weinmann2006} find that (${\sim} 40-50\%$) of the central galaxies in galaxy groups are star-forming late-type galaxies.
In fact,  a significant minority of galaxies has been shown to have disky morphology even at the highest stellar mass (M$_*$ > 10$^{11.4}$ M$_\odot$) (e.g.
\citealt{conselice2006early,ogle2016superluminous,ogle2019break}) and even early-type BGGs have been shown to contain non-trivial amount of cold gas \citep{werner2014origin,o2018cold}.
The presence of cold gas and gaseous disks, rejuvenations, and the ongoing star formation all indicate that BGGs must be receiving influx of gas from their surroundings every so often.  In the present paper, we focus on the reservoir of gas cocooning the {\sc Romulus} BGGs (hereafter, the circumgalactic medium or the CGM) and seek to identify the different modes of gas influx.  In the process, we also investigate the nature of the CGM and the origin of structure therein.  As it turns out, all of these are inter-linked. 

The nature and the dynamics of the CGM and the impact of these on galaxy formation is a highly topical subject.  Observations of CGM surrounding BGGs and giant elliptical galaxies find that this reservoir comprises gas with a broad range of kinematics, 
ionization states, temperatures, densities, and phases \citep[e.g.][]{werner2014origin,werk2016cos,lakhchaura2018thermodynamic,o2018cold,zahedy2019characterizing,olivares2022gas}.
Among the many riddles linked to this rich diversity, the origin of cold ($\sim 10^4$ K) gas co-existing alongside warm-hot ($\sim 10^6-10^7$ K) gas in the CGM of quiescent galaxies is particularly puzzling.  Possible explanations of the phenomenon run the gamut from (i) debris from ram pressure stripping and tidal disruption of satellite galaxies (e.g. \citealt{murakami1999interaction,mccarthy2008ram,sun2009spectacular,franchetto2021evidence, jung2022massive}); (ii) byproducts of cooling due to mixing between the diffuse CGM and the multiphase stellar and/or AGN-driven galactic outflows \citep{Huang2020,huang2020new,schneider2020outflowmix,Rennehan2021}, as well as mixing between the diffuse CGM and
the interstellar gas dredged (uplifted) out of the galaxy by active galactic nucleus (AGN) jets  (e.g. \citealt{revaz2008formation,pope2010mass,scannapieco2015launching,
schneider2017hydrodynamical,qiu2020formation});  
to (iii) the byproducts of induced positive 
density perturbations within the CGM that result in localized regions of rapidly cooling gas which condense out of the general medium and form in-situ cool clouds (e.g. \citealt{maller2004multiphase,sharma2010thermal,
sharma2012thermal,choudhury2016cold,prasad2015cool,prasad2017agn,voit2015precipitation,li2015cooling}).\footnote{
In literature, the phenomenology is often associated with labels ``condensations'' and ``precipitation''.   In this paper, ``condensations'' refers to cool-cold clouds forming in-situ out of localized regions of rapidly cooling density perturbations.  Precipitation refers to cold clouds that fall or rain down onto the central galaxy.}  The key to sorting through these various mechanisms, and the way in which they impact the evolution of the galaxies  is to undertake a detailed study of the baryon cycles and the physics underlying these cycles in these systems \citep[cf.~][and references therein]{DonahueVoit2022baryons,FaucherOh2023CGMphys}. 

For systems that assemble hierarchically over the cosmic timescales,
one way of quantifying the importance of the different origin mechanisms and more generally, of acquiring better insight into the dynamics of the CGM, is to track its evolution in self-consistent, cosmological hydrodynamic simulations of galaxy evolution.  However, investigating the CGM using cosmological simulations is very challenging because most current cosmological simulations typically under-resolve lower density structures, like the CGM \citep{hummels2019impact}.  This is due to two main constraints:  (i) an effective ceiling on the maximum resolution due to practical considerations (e.g., the computational costs in terms of time and resources), and (ii) the intentional design of the codes to prioritize and direct computational resources towards high density structures.   

There are, however, concerted ongoing efforts to improve the treatment of the CGM.   One promising approach involves code redesign \citep[cf.][]{hummels2019impact,suresh2019zooming,van2019cosmological,peeples2019figuring} that allows the mass resolution in a halo’s circumgalactic environment to be boosted above the halo’s base resolution.  The other approach leverages more efficient codes, faster computers and greater computing resources to push to higher resolutions overall \citep[cf.][]{ tremmel2019introducing,fiacconi2017young,hafen2019origins,nelson2019tng50,fielding2020first,tamfal2022dawn}.   Our {\sc Romulus} suite of simulations fall in the latter category.   In fact, most of the studies cited above that follow the evolution of cosmologically realistic halos rely on higher resolution globally, the only exceptions being \citet{suresh2019zooming}, \citet{van2019cosmological} and \citet{peeples2019figuring}.

To be sure, both approaches have limitations.  For one, neither class of simulations have achieved convergence.   Nonetheless, the effort represents significant improvement.  Also, we note that at present the computational expenses involved typically limit the size of the halos that can be investigated, or restrict the runs from being continued past some time point, or require some other compromises.  For example, all except \citet{nelson2020resolving} and the {\sc Romulus} analyses limit themselves to halos of mass M$_\mathrm{vir} \sim 10^{12}$ M$_\mathrm{\odot}$ or lower.

Here, we expand on prior {\sc Romulus} studies \citep{tremmel2017romulus,tremmel2019introducing,butsky2019ultraviolet,chadayammuri2020fountains,jung2022massive} to investigate the origin and the dynamics of the CGM in the radial range $\rm 0.1 \leq R/R_\mathrm{500} \leq 1$ around the central galaxies in group-scale halos with present-day ($z=0$) masses in the range  $2.3 \times 10^{12}$  M$_{\rm \odot} \lsim M_\mathrm{200}\lsim 1 \times 10^{14}\;$ M$_{\rm \odot}$.  When we speak of the CGM, this is the gas that we are referring to --- unless we explicitly say otherwise.
We exclude the gas inside 0.1R$_\mathrm{500}$ from current study because gas dynamics in this ``inner CGM'' is considerably more complex. For one, it undergoes repeated heating and cooling episodes which give rise to variations in the central temperature, cooling times, gas velocities, etc.  We will explore the evolution of the ``inner CGM'' in a follow-up paper. This paper is organized as follows: In Section \ref{method}, we explain our analysis methodology and present some preliminary results. In Section \ref{origin}, we discuss the nature of the CGM and investigate its origin. We then explore the dynamics of the multiphase CGM in Section \ref{thermodynamics}. We summarize and discuss our results in Section \ref{discussion} and give final concluding remarks in Section \ref{conclusion}.

\section{Analysis Methods and Initial Findings}\label{method}
\begin{table*}
\fontsize{9}{11}\selectfont
\centering
\begin{tabular}{|l|p{1.1cm}|p{0.6cm}|p{1.25cm}|p{1.25cm}|p{1.15cm}|p{1.15cm}|p{0.7cm}|p{0.7cm}|p{0.7cm}|p{1cm}|p{0.7cm}}
 \hline
 \hline
 \\
 Halo & M$_\mathrm{200}\rvert_{z=0}$  
 & $z_\mathrm{a}$ & M$_\mathrm{200}$ & M$_\mathrm{500}$ & M$_{\mathrm{*},
\mathrm{R}_\mathrm{gal}}$  & M$_{\mathrm{*}, 50\; \mathrm{kpc}}$  & R$_\mathrm{200}$  & R$_\mathrm{500}$  & R$_\mathrm{gal}$  \\
    ID     & [M$_\mathrm{\odot}$] &   & [M$_\mathrm{\odot}$] & [M$_\mathrm{\odot}$] & [M$_\mathrm{\odot}$] & [M$_\mathrm{\odot}$]  &   [kpc] & [kpc]  & [kpc] \\
 \hline
 \\
 C  & 1.1e+14 & 0.70 &5.9e+13 & 4.6e+13 & 8.0e+11 & 9.2e+11 & 629.8 & 428.5 & 23.0   \\
 52024 &1.8e+13& 0.29  & 1.3e+13 & 1.0e+13 & 2.0e+11 & 2.5e+11 & 451.7 & 304.6 & 15.2  \\
 G1 &1.4e+13 & 0.25 & 1.6e+13 & 1.1e+13 & 2.6e+11 & 3.3e+11 & 488.9 & 317.7 & 15.9  \\
 49510 & 1.2e+13&  0.36  &1.1e+13 & 8.2e+12 & 1.7e+11 & 2.2e+11 & 417.5 & 276.6& 12.1  \\
 99966 & 6.3e+12 & 0.31 &5.9e+12 & 3.9e+12 & 1.1e+11 & 1.3e+11 & 342.2 & 219.8& 12.1 \\
 38182 & 4.4e+12 & 0.44 &4.0e+12 & 2.8e+12 & 1.0e+11 & 1.1e+11 & 286.3 & 187.2& 11.0  \\
 91655 & 2.3e+12 & 0.26 &2.1e+12 & 1.7e+12 & 1.1e+11 & 1.4e+11 & 247.9 & 171.3 & 13.3 \\
 77876 & 2.3e+12 & 0.25  &2.1e+12 & 1.8e+12 & 1.0e+11 & 1.3e+11 & 251.2 & 174.2 & 11.0 \\
 \\
\hline
\end{tabular}
\caption{Properties of the eight {\sc Romulus} halos that we use as foil for the discussion in this study. Halos are ordered by halo mass M$_\mathrm{200}$ at z = 0.  $z_\mathrm{a}$ is the redshift at which the halos were analyzed.  M$_\mathrm{200}$ (M$_\mathrm{500})$ is the the total mass at $z_\mathrm{a}$ within R$_\mathrm{200}$ (R$_\mathrm{500})$.  M$_\mathrm{*}$(< R$_\mathrm{gal})$ is the total stellar mass contained within the central galaxy's radius R$_\mathrm{gal}$ (see text for definition of R$_\mathrm{gal}$) and M$_\mathrm{*}$(< 50 kpc) is stellar mass within 50 kpc of the halo center at $z_\mathrm{a}$.}

\label{table:table1}
\end{table*}

We use three of the {\sc Romulus} simulations: {\sc Romulus25}, which is a ($25\,{\rm Mpc})^3$ cosmological volume simulation, as well as {\sc RomulusC} and {\sc RomulusG1}, which are two zoom simulations of individual group-scale systems.  All three simulations share the same background cosmology\footnote{The background cosmology corresponds to a $\Lambda$CDM universe with cosmological parameters consistent with \citet{Planck_2016} results: $\Omega_{\rm m} = 0.309$, $\Omega_{\rm \Lambda} = 0.691$, $\Omega_{\rm b} = 0.0486$, $\rm H_{\rm 0} = 67.8\, {\rm km}\;{\rm s}^{-1} \rm Mpc^{-1}$, and $\sigma_{\rm 8} = 0.82$.}, sub-grid physics and {\sc Romulus} galaxy formation model \citep{tremmel2017romulus,tremmel2019introducing,butsky2019ultraviolet,chadayammuri2020fountains,jung2022massive}.
Also, all three simulations are run using Tree+Smoothed Particle Hydrodynamics (Tree+SPH) code {\sc CHaNGa} \citep{menon2015adaptive,wadsley2017gasoline2}. 

The details about the {\sc Romulus} simulations, including a thorough discussion of the hydrodynamics code and the galaxy formation model used, the sub-grid physics incorporated therein, the various modeling choices adopted, as well as the simulations' unique features, have been described extensively in a number of published papers.  In the interest of brevity, we do not repeat this account here and instead, refer interested readers to \citet{tremmel2015off,tremmel2017romulus,tremmel2019introducing,tremmel2020formation,sanchez2019not,butsky2019ultraviolet,chadayammuri2020fountains}; and \citet{jung2022massive}.  The latter especially offers a concise yet complete summary.  Here, instead, we concentrate on providing a clear description of the approach we take in analyzing the simulations.

 There is, however, one aspect that is important to highlight: The resolution of all the groups extracted from {\sc Romulus} simulations is the same regardless of whether they are drawn from zoom or cosmological simulations. They are all characterized by a Plummer equivalent gravitational force softening of 250 pc (or 350 pc spline kernel), a maximum SPH resolution of 70 pc, and  gas and dark matter particle masses of $2.12 \times 10^{5}\; \rm M_{\odot}$  and  $3.39 \times 10^{5}\; \rm M_{\odot}$, respectively. 
 
 Together, the {\sc Romulus} simulations yield a sample numbering in a few tens of halos in the mass range of interest. These are among the highest resolution group-scale halos simulated using cosmologically realistic initial conditions.

\subsection{Halo catalogue and substructure definition}

The output files for all the {\sc Romulus} simulations discussed in this study were processed in an identical manner: The halos and subhalos  were extracted and processed using the Amiga Halo Finder \citep[AHF]{knebe2008relation,knollmann2009ahf}, and tracked across time with TANGOS \citep{pontzen2018tangos}.  

The halos and subhalos exist in a nested hierarchy, with the halos being the top-level structures and the subhalos embedded therein. AHF finds these structures by locating peaks in an adaptively smoothed density field, identifying all particles (dark matter, gas, stars, and black holes) that are gravitationally bound to each peak, and then proceeding up the hierarchy to find successively larger structures.  Once the halos are identified, their centers are found by applying the shrinking sphere approach \citep{power2003inner} to the distribution of bound particles associated with each of the halos. 

The halo masses $(\rm M_\Delta)$ are determined by constructing a sphere of radius $\rm R_\Delta$ about each of the halo centers such that the mean interior density is $\left\langle{\rho_\mathrm{m,\Delta}}(z)\right\rangle = \Delta \cdot \rho_\mathrm{crit}(z)$ where $\rho_\mathrm{crit} = \rm 3H^2(z)/8\pi G$ is the critical cosmological density and $\Delta$ is a constant.  \citep[see, for example,][]{babul2002physical}. Throughout this study, we refer to ($\rm M_\mathrm{200} , R_\mathrm{200}$) and ($\rm M_\mathrm{500} , R_\mathrm{500}$)  which correspond to $\Delta = 200$ and $\Delta = 500$ respectively. For the assumed cosmology in {\sc Romulus} simulations $\rm M_\mathrm{500}/M_\mathrm{200} \sim 0.7 $ and $\rm R_\mathrm{500}/R_\mathrm{200} \sim 0.68 $. 

The above prescription is only appropriate for the halos.  In the case of subhaloes, AHF tracks the  local density profile as a function of distance from the peak center. At some point, the external gravitational field becomes dominant and correspondingly, the behaviour of the density profile changes.   
The distance from the peak where this happens marks the size of the subhalo, and the mass enclosed is recorded as the subhalo's mass.

As noted in \S \ref{Introduction}, in this study we are interested in the CGM surrounding the massive central galaxies in group-scale halos.   Consequently, we restrict ourselves to halos in our simulations with $z=0$ masses $\rm M_{\rm 200} \geq 2 \times 10^{12}\; \rm M_\mathrm{\odot}$ \citep[cf][]{jung2022massive}.

\subsection{Central galaxy selection criterion}\label{BGG}

Next, we are specifically interested in the evolution of the quiescent CGM, that is the dynamics of the CGM in the absence of shocks and other disturbances due to major mergers and massive interlopers \citep[see, for example,][]{poole2006mergingI}.  We therefore examine each of the halos of interest and locate a 5 Gyrs window over the redshift range $0 \leq z \leq 1.5$ during which the volume inside $\rm R_\mathrm{500}$ is undisturbed by a major (1:10 or larger) merger.   We set our epoch of analysis (hereafter, $z_{\rm a}$) at two Gyrs after the start of the window. Our analysis of the gas dynamics in group-scale halos that undergo major mergers shows that it rarely takes longer than two Gyrs for the gas inside $\rm R_\mathrm{500}$ to relax.  In order to identify the origins and the fate of the CGM gas at the time of analysis, $z_{\rm a}$, we consider the evolution of the gas starting from 5 Gyrs prior (we explain the choice of 5 Gyrs in \S\ref{subhalosection}) to $z_{\rm a}$ to 3 Gyrs after.

Since the halos are generally dynamically relaxed during the period of study, the massive galaxies in the halos are always located at the halo center.  We determine the galaxies' baryonic properties by looking at the gas, stellar and black hole particles within spheres of different sizes centered on the halo centers.  To further restrict ourselves to massive central galaxies (hereafter, BGGs), we only investigate systems that host central galaxies with $ \rm M_{\rm *}(< \rm R_\mathrm{\rm gal})  \geq 10^{11}\; \rm M_{\rm \odot}$, where $\rm R_\mathrm{\rm gal}$ is the galaxy's radius. We follow \citet{hafen2019origins} and define a central galaxy's radius as $\rm R_\mathrm{gal} = 4 \rm R_\mathrm{*,0.5}$, where $\rm R_{*,0.5}$ is the half mass radius for all star particles within 0.15{$\rm R_\mathrm{200}$}.

There are 19 such systems across all of the {\sc Romulus} simulations and although we have analyzed all 19, we have selected
eight for detailed investigation and as foil for our discussion.
The basic characteristics of these eight halos are listed in Table\ref{table:table1}.  We emphasize that these should be regarded as illustrative. They were chosen because their halo and BGG masses, as well as the number and properties of the subhalos orbiting inside $\rm R_\mathrm{500}(z_{\rm a})$ sample the range spanned by the full sample, and because their CGM properties reflect the diversity present in our sample.  For example, we wanted to ensure that the sample included both cool core and non-cool core groups. We emphasize that the findings and behaviours that we highlight are general and observed in all the halos. 

For completeness, we point out that the cool core/non-cool core classification criterion for observed groups is \emph{not} the same as that for clusters.  For the latter, it is based on entropy or the cooling time of the intracluster medium in the cluster core.  Groups, however, are classified on the basis of  their observed temperature profiles:  Those that exhibit central temperature drop are designated ``cool core (CC)'' while those with flat or centrally peaked temperature profiles are labelled ``non-cool core (NCC)''.  We refer interested readers to \citet{o2017complete} for further details.   The classification of {\sc Romulus} groups is discussed in \citet{jung2022massive} and their temperature profiles are shown in Fig.~12 of that paper.

   \begin{figure*}
   \centering
   \includegraphics[width=0.9\linewidth]{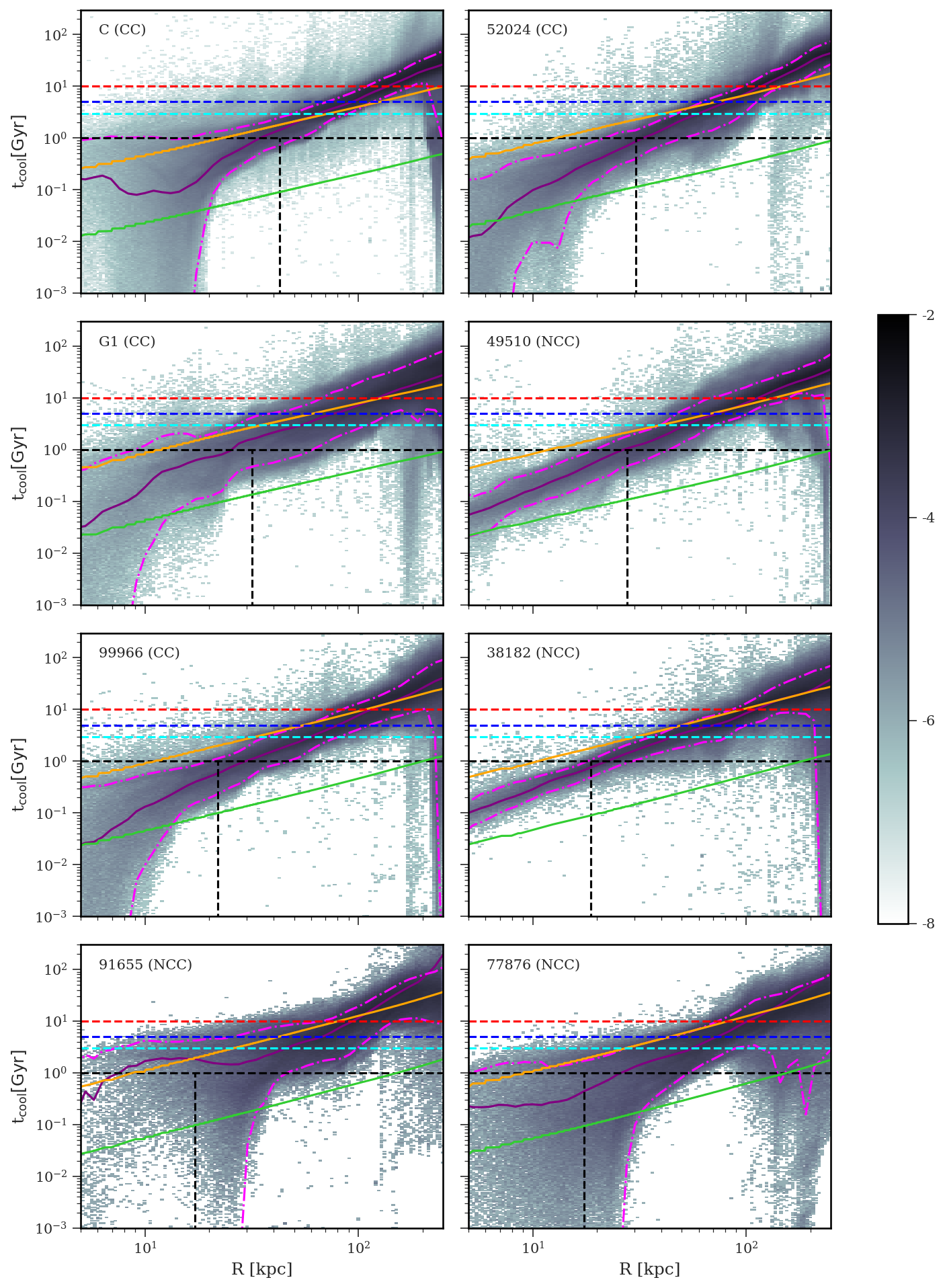}
      \caption{The distribution of $t_\mathrm{cool} $ for gas within $\rm R_\mathrm{500}$ in eight {\sc Romulus} halos used in this study at $z_\mathrm{a}$. The grey pixels show $t_\mathrm{cool}$ versus the radius of \emph{all} gas. The pixels' contrast is determined by the logarithm of the ratio of gas particles in each pixel to the total gas particles within $\rm R_\mathrm{500}$ of each halo. The purple line in each panel is the median $t_\mathrm{cool}$ for the hot (T > 10$^6$ K) gas. The dashed horizontal lines corresponds to $t_\mathrm{cool}$ = 10 (red), = 5 (blue), = 3 (cyan) and = 1 (black).The magenta dotted-dashed lines show the upper and lower values of $t_\mathrm{cool}$ bounding 90$\%$ of non-ISM gas.  Black vertical line shows r = 0.1 $\rm R_\mathrm{500}$ for each halo. As discussed in the text, we designate the gas at  $0.1 \leq \rm R/R_\mathrm{500} \leq 1$ as the CGM.  The orange line shows $t_\mathrm{cool}$/t$_\mathrm{ff}$ = 20 and green line shows $t_\mathrm{cool}$/t$_\mathrm{ff}$ = 1.  The significance of these thresholds is discussed in the text. 
      }
      \label{tcool}
   \end{figure*}

\subsection{The circumgalactic gas surrounding the BGGs}\label{CGM}

The main focus of this study is to determine where the gas in the CGM around massive galaxies at the time of analysis, $z_{\rm a}$, has come from, to identify the different types of structures present, and  ascertain the fate of these CGM subcomponents.  We will do this via particle tracking analysis applied to the individual gas resolution elements of the CGM.  Particle tracking leverages the fact that Lagrangian hydrodynamic codes, like {\sc CHaNGa}, provide access to the full time history of these resolution elements and the fact that mass exchange between gas elements is not permitted.  It has previously been used to study gas inflow onto galaxies and their halos, as well as the impact of galactic winds; it has also been used to study the gas and metal content of the intragroup medium and most recently, the CGM around Milky Way-like and lower mass halos
\citep[cf.~][and references therein]{keres2005accretion,oppen2010wind,liang2016igrm,angles2017baryons,hafen2019origins,esmerian2021thermal}.

An assessment of the gas properties (e.g., temperature, entropy, cooling time, etc.) surrounding the BGGs in the {\sc Romulus} groups shows that at any given radius, these typically span a broad distribution. 
This is illustrated in Fig. \ref{tcool}, where we show the distribution of the local gas cooling time at $z_\mathrm{a}$, as a function of halocentric distance, in our eight illustrative halos. 
Here,  $t_\mathrm{cool}  \equiv E_\mathrm{thermal}/[n_e^2 \Lambda]$, where $\Lambda$ is the radiative cooling function and $n_e$ is electron number density.

The grey shading in each of the panels illustrates the distribution of \emph{all} the gas inside R$_\mathrm{500}$ of the identified halos, including the cool, potentially star-forming gas (hereafter, the interstellar medium or the ISM) within the BGG.  Operationally, we divided the panels into equal-sized pixels, binned the gas particles in these pixels, and converted the resulting count into a fraction relative to the total number of gas particles within R$_\mathrm{500}$.  The contrast of the shading is determined by the logarithm of the fraction: larger the fraction, darker the shading.
Additionally, the horizontal dashed lines in each of the panels correspond to cooling timescales of 10 Gyrs (red), 5 Gyrs (blue), 3 Gyrs (cyan), and 1 Gyr (black). The purple curve is the median $t_\mathrm{cool}$ for the hot (T > 10$^6$ K) gas; the vertical black dashed line corresponds to 0.1R$_\mathrm{500}$; and the two magenta dot-dashed lines show the upper and lower values of  $t_\mathrm{cool}$ bounding 90\% of the non-ISM gas.\footnote{Specifically, we exclude any gas that is bound to central and satellite galaxies \emph{and} whose temperature is less than $ 5\times 10^4 K $ and hydrogen number density greater than 
$0.1 \, \mathrm{cm}^{-3}$.}  As for the orange and green diagonal lines, we will discuss these in \S \ref{fullCGM}.

 \subsubsection{The CGM}\label{fullCGM}

The panels in Fig. \ref{tcool} show several features worth noting. We first start by considering all the gas in the radial range $0.1  \leq \rm R/R_\mathrm{500} \leq 1 $.  We find that no more than 10\% (20\%)  of the gas has cooling time of less than 1 Gyr (3 Gyrs).  Generally, the non-ISM CGM gas in all halos is distributed in a band about the median hot gas $t_\mathrm{cool}$ curve.  Still, even if we set aside the cool dense ISM-like gas in the satellite and the central galaxies \emph{and} further exclude the gas in the two tails of the distribution,  the remaining 90\% (bracketed by the two magenta dot-dashed lines) includes gas whose cooling times range from $\sim 0.3-0.4$ to $\sim 2-3$ times the median \emph{at the same radius}. Moreover, we also find occasional deep downward extensions plunging cooling times as much as $10^{-3}\times$, or even $10^{-4}\times$
lower, as well as upward extensions from the median band reaching $\sim$100$\times$ larger, than the hot gas median.  Such large dynamic ranges in $t_\mathrm{cool}$ has also been reported by \citet{esmerian2021thermal} and \citet{nelson2020resolving} in simulations of Milky Way-mass (FIRE-2) as well as group-scale ($M_{200}\sim 10^{13.5}\;M_{\rm \odot}$) halos (TNG50), respectively. As for the origin of the dips and peaks, we will discuss these further in greater detail in \S \ref{origin} \& \S \ref{thermodynamics}.  Here we simply note that the former are due to gas recently stripped from the satellites while the latter are due to shock-heated gas.

Secondly, although we have explicitly indicated that we will focus on gas in the radial range $0.1  \leq \rm R/R_\mathrm{500} \leq 1 $ around BGGs, we cannot ignore the fact that in many of the halos,  bulk of the gas --- and this includes the non-ISM gas --- inside 0.1R$_\mathrm{500}$  has short ($<$ 1 Gyr) cooling time.  Consequently during periods of low AGN activity, one expects the emergence of a global cooling flow. \citet{chadayammuri2020fountains} noted this trend in {\sc Romulus C} system (see their Fig.~7 and 8), and we see it in \emph{all} other {\sc Romulus} groups as well (see \S \ref{coolingflow}). Inevitably, these flows are disrupted by AGN outburts (e.g.~see panel for halo 91655 in Fig. \ref{tcool}) but when present, they extend beyond 0.1R$_\mathrm{500}$ and impact the dynamics of the gas of interest (cf. \S\ref{coolingflow}).

Next, we turn to the solid orange diagonal lines.  This line corresponds to $t_\mathrm{cool} / t_\mathrm{ff}=20$, where $t_\mathrm{ff}  \equiv \sqrt{2r/g(r)}$ is the free-fall time at radius $r$ and $g(r) = GM(<r) / r^2$.  We consider this quantity because linear stability analyses and experiments with idealized simulations find that small-amplitude density perturbations in stratified diffuse CGM in global thermal balance are thermally unstable and susceptible to condensation whenever the median $t_\mathrm{cool} / t_\mathrm{ff}$ drops below 10. 
\footnote{While the gas is formally linearly thermally unstable for all $t_\mathrm{cool} / t_\mathrm{ff}$ \citep{choudhury2016cold}, nonlinearly, it becomes multiphase when the median ratio is lower than 10. In this work, we refer to the gas prone to condensation as ``condensation-susceptible'' gas.}
Condensing perturbations will form cool-cold clouds, giving rise to multiphase structure in the CGM (cf.\citealt{sharma2010thermal,sharma2012thermal,
mccourt2012thermal,meece2015condensations,choudhury2016cold}). And, while there is observational support for this thermal instability picture from, for example, studies of the galaxy cluster cores, some of these same observational studies suggest a higher $t_\mathrm{cool} / t_\mathrm{ff}$ value for the onset of the instability (e.g., $\sim 20$; \citealt{hogan2017onset,pulido2018origin,Babyk2018}).  Recently,  a theoretical reanalysis by
\citet{choudhury2016cold} suggests that this value
depends on the detailed shape of the gravitational potential well \citep[due, for example, to the central galaxy within the halo; see also][]{voitdonahue2015accept,prasad2018cool} and can be as much as a factor of 2 higher in realistic group/cluster cores. 
Guided by these findings, we adopt  $t_\mathrm{cool}/t_\mathrm{ff} = 20$  (solid orange line) as our threshold (in \S\ref{threshold}, we discuss how our findings change if our threshold is raised to 30 or lowered to 10.)  
An examination of the panels in Fig. \ref{tcool} shows that in all except halo 91655 (which is undergoing active AGN heating), the purple curve lies below the solid orange curve (or equivalently, $(t_\mathrm{cool} / t_\mathrm{ff})_\mathrm{med} \leq 20$) out to $\sim 60$ kpc.  In other words, gas density perturbations within the central $\sim 60$ kpc of these group-scale halos ought to be prone to condensation.

The green solid line corresponds to $t_\mathrm{cool} / t_\mathrm{ff}=1$. \citet{voit2021graphical} suggests that once the perturbations' local $t_\mathrm{cool} / t_\mathrm{ff}$ drops below $t_\mathrm{cool} / t_\mathrm{ff}$ = 1, they will condense and form cool-cold clouds. We will discuss the distinction between local and global values of $t_\mathrm{cool} / t_\mathrm{ff}$ in \S \ref{tuCGM}.

\begin{figure*}
   \centering
   \includegraphics[width=0.89\linewidth]{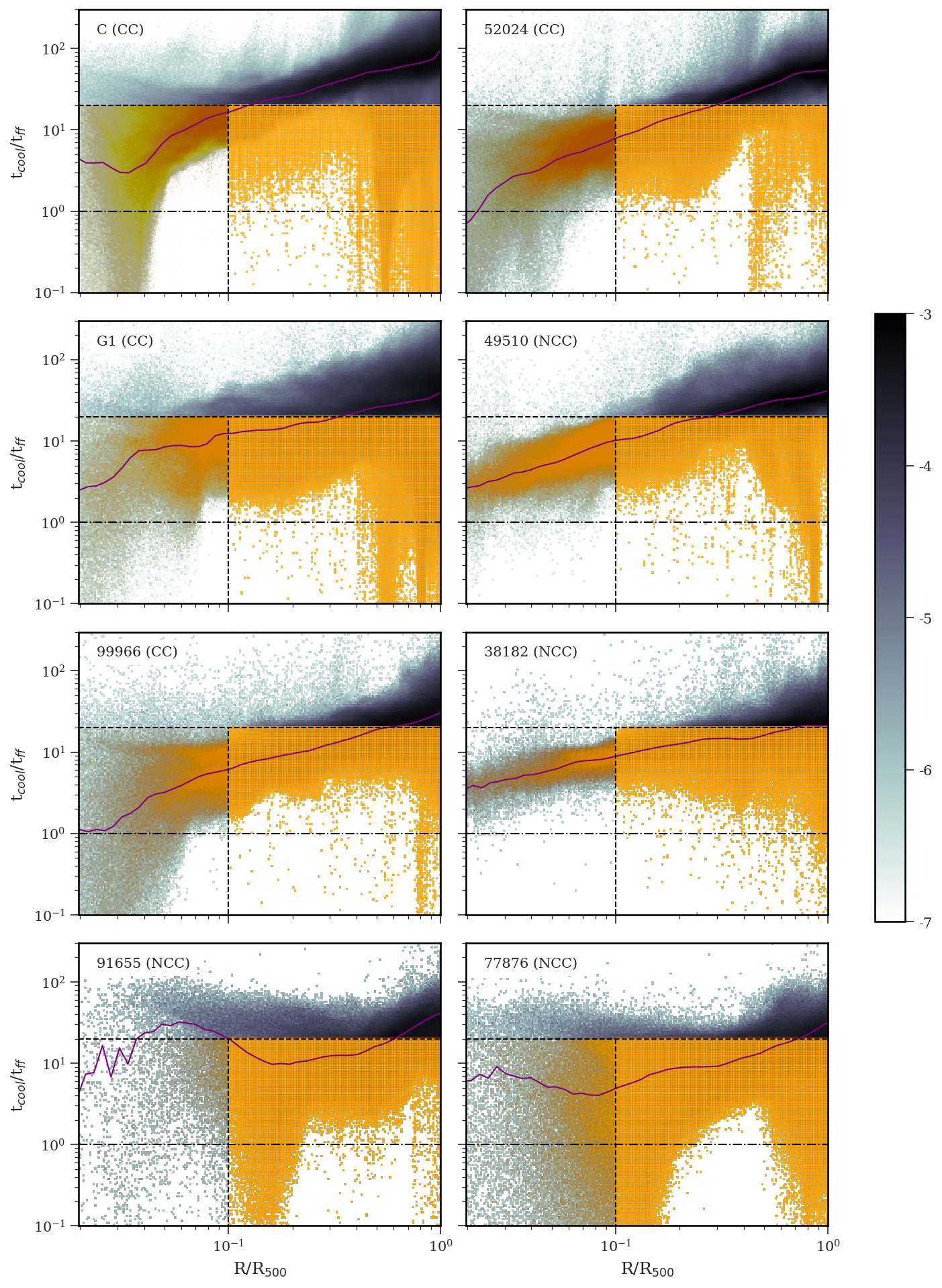}
      \caption{The distribution of $t_\mathrm{cool} / t_\mathrm{ff}$ for gas within $\rm R_\mathrm{500}$ in {\sc Romulus}
      groups under consideration at $z_\mathrm{a}$. The grey pixels show $t_\mathrm{cool}$/$t_\mathrm{ff}$ versus the radius of \emph{all} gas.  The radius is scaled by characteristic radius (R$_\mathrm{500}$). The grey shading has the same meaning as Fig.\ref{tcool}. The purple line is the median $t_\mathrm{cool}$/t$_\mathrm{ff}$ for the hot (T > 10$^6$ K) gas. The dashed black vertical line shows r = 0.1 $\rm R_\mathrm{500}$. As discussed in the text, we designate the gas at  $0.1 \leq \rm R/R_\mathrm{500} \leq 1$ as the CGM.  The dashed, black, horizontal line shows $t_\mathrm{cool}$/t$_\mathrm{ff}$ = 20 and dash-dotted horizontal line shows $t_\mathrm{cool}$/t$_\mathrm{ff}$ = 1.  The significance of these thresholds is discussed in the text.  The orange shading shows the distribution of ``condensation-susceptible'' CGM. We define this as gas between  $0.1 \leq \rm R/ R_\mathrm{500} \leq 1$ with $t_\mathrm{cool} / t_\mathrm{ff} < 20$ as well as gas between $0.02 < \rm R/R_\mathrm{500} < 0.1 $ that has $t_\mathrm{cool} / t_\mathrm{ff} < 20$ at $z_\mathrm{a}$ \emph{and} which entered this domain from beyond 0.1$\rm R_\mathrm{500}$ in the 5 Gyrs prior to the time of analysis.  We explicitly exclude ISM gas. 
      }
      \label{Fig2}
   \end{figure*}

In Fig.~\ref{Fig2}, we show the distribution of the local $t_\mathrm{cool} / t_\mathrm{ff}$ at $z_\mathrm{a}$ for \emph{all} the gas inside R$_\mathrm{500}$, including the cool, potentially star-forming ISM gas, in halos under consideration. The grey shading has the same meaning as in Fig. \ref{tcool}; the purple curve is the median $t_\mathrm{cool} / t_\mathrm{ff}$ profile for the hot (T > $10^6$ K) gas; and the dashed and dash-dotted horizontal lines corresponds to the $t_\mathrm{cool} / t_\mathrm{ff} = 20$ and $t_\mathrm{cool} / t_\mathrm{ff} = 1$, respectively. Additionally, we have shaded some of the gas in the panels translucent orange.  We discuss this component in detail in \S \ref{tuCGM} below but first we examine the properties of the CGM gas (i.e.~the gas in the radial range $0.1  \leq \rm R/R_\mathrm{500} \leq 1 $).

Examining the distribution of $t_\mathrm{cool} / t_\mathrm{ff}$, we frequently find a large spread in the same radial shell.   Specifically, $t_\mathrm{cool} / t_\mathrm{ff}$ extends to values as low as $\sim$ 0.1, and in some instances down to $10^{-3}$ or even $10^{-5}$, and as high as $\sim$ 300. This again highlights 
the multiphase nature of the CGM.  \citet{esmerian2021thermal} and \citet{nelson2020resolving} also find similarly large dynamic range in $t_\mathrm{cool} / t_\mathrm{ff}$ in their simulations.

\begin{figure*}
   \centering
   \includegraphics[width=\linewidth]{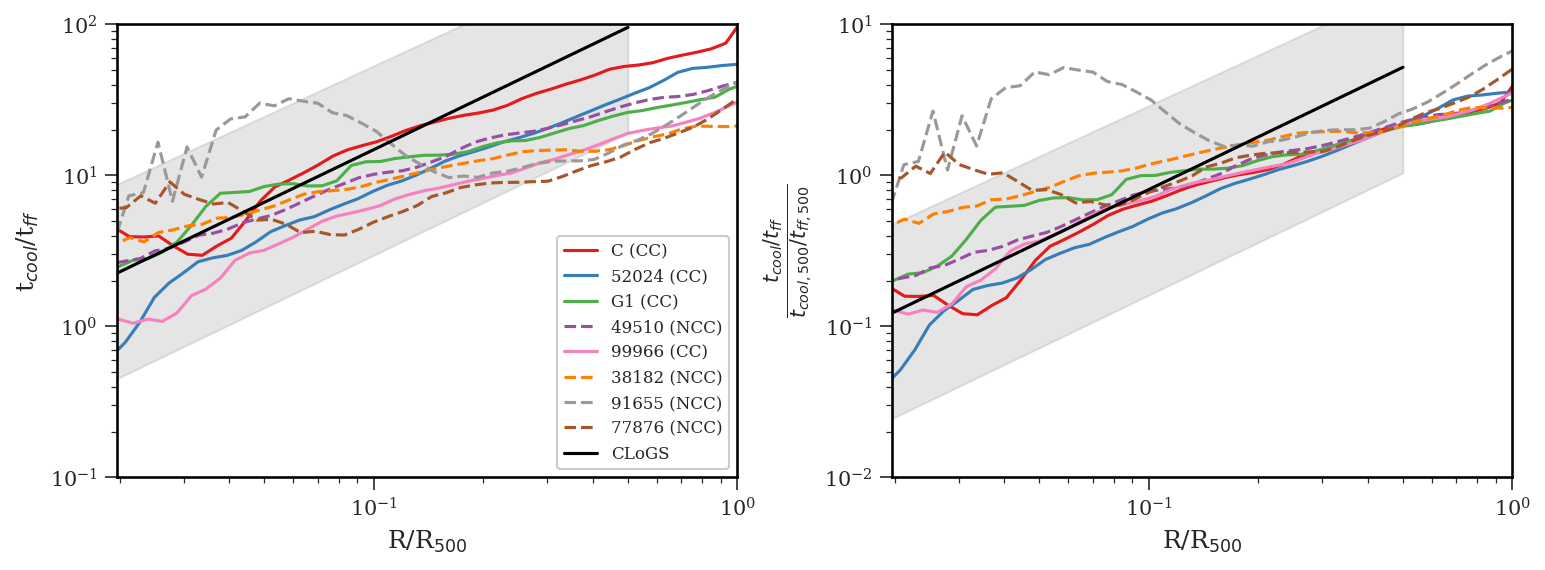}
      \caption{Median $t_\mathrm{cool}$/t$_\mathrm{ff}$ of hot (T > 10$^6$ K) gas for the eight {\sc Romulus} halos vs radius. The left panel juxtaposes all median lines from Fig. \ref{Fig2}. We show the median line for CC groups as solid lines, and NCC as dashed lines. The right panel shows the same $t_\mathrm{cool}$/t$_\mathrm{ff}$ profiles, scaled by t$_\mathrm{cool,500}$/t$_\mathrm{ff,500}$ following \citet{prasad2020cool}. Black solid line is the median $t_\mathrm{cool}$/t$_\mathrm{ff}$ for observed CLoGS groups \citep{o2017complete} and the grey band shows the observed spread around the CLoGS median.  
      }
      \label{clogs}
   \end{figure*}

Next, we consider the median {\sc Romulus} $t_\mathrm{cool} / t_\mathrm{ff}$ profiles (purple curve in each panel) for the hot (T > $10^6$ K) gas --- see also, the left panel of Fig. \ref{clogs}, where all eight are juxtaposed for easy comparison. These profiles can be directly compared to the $t_\mathrm{cool} / t_\mathrm{ff}$ profiles in Fig. 11 of \citet{nelson2020resolving}, which are for the hot gas in TNG50 groups of comparable masses to our {\sc Romulus} groups.  
In terms of resolution, the TNG50 simulation is similar to {\sc Romulus}: As described by \citet{pillepich2019first}, TNG50 has gas and dark matter element mass of  $8.5 \times 10^{4}\; \rm M_{\odot}$ and $4.5 \times 10^{5}\; \rm M_{\odot}$, respectively [versus gas and dark matter particle masses in {\sc Romulus} of $2.12 \times 10^{5}\; \rm M_{\odot}$  and  $3.39 \times 10^{5}\; \rm M_{\odot}$, respectively];  a Plummer equivalent gravitational force softening of 288 pc [versus 250 pc in {\sc Romulus}]; and a minimum adaptive gas gravitational softening of 72 pc [versus a minimum SPH resolution of 70 pc].

Nevertheless, the results are strikingly different: All except two of our median profiles increase monotonically with radius, power-law-like, for  $r \gsim 10\; {\rm kpc}$ ($r \gsim 0.05$R$_\mathrm{500}$). Even the median profiles for halos 91655 and 77876, which feature a ``bump''\footnote{The ``bump'' is due to ongoing AGN outbursts (see discussion and figure in \S \ref{origin} for further details). Similar central bump-like features are present in the observationally determined $t_\mathrm{cool}$ profiles of the CLoGS galaxy groups with jets (cf. Fig. 5 of \citealt{o2017complete}.)} within the central $\sim 0.1-0.2\ \rm R/R_\mathrm{500}$, fall in with the rest and then rise towards larger radii.  In contrast, the TNG50 profiles rise, have a maximum at $15-40\;{\rm kpc}$, and then decrease towards larger radii.

The shape of the {\sc Romulus} profiles are in agreement with the observed $t_\mathrm{cool} / t_\mathrm{ff}$ profiles in galaxy groups \citep[cf.][]{o2017complete}, which also increase with radius for $\rm r > 0.1 R_\mathrm{500} $.  This is illustrated in  Fig. \ref{clogs} by the solid black line, which corresponds to the observationally determined median $t_\mathrm{cool} / t_\mathrm{ff}$ profile for the subset of the CLoGS galaxy groups \citep{o2017complete} with available X-ray data, and the grey band, which illustrates the spread about the observed median.  
In the left panel, it is clear that the CLoGS profiles, like the {\sc Romulus} profiles, increase with distance from the group center. However, the plot also gives the impression that the observed profiles have a slightly higher normalization.  This offset is 
due to  the fact that the {\sc Romulus} halos span a larger range in $z_\mathrm{a}$ and halo mass than the CLoGS sample.  In particular, the {\sc Romulus} sample has more lower mass systems than CLoGS.   We follow the scaling procedure described by \citet{prasad2020cool} to account for these variations and the results are shown in the right panel of Fig. \ref{clogs}.  The use of scaled $t_\mathrm{cool} / t_\mathrm{ff}$ leads to the narrowing of the scatter between the {\sc Romulus} profiles,  especially for R > 0.1R$_\mathrm{500}$, and an improved agreement with the normalization of the CLoGS curve. Rescaling the CLoGS median/band does not have much impact because the groups are all nearby ($z < 0.02$) and span a narrow range in mass and redshift.

Returning to the \citet{nelson2020resolving} results, we cannot discern whether the differences in the shape of their and our profiles are due to the baryon mass resolution of TNG50 being slightly better than that of {\sc Romulus};
due to the differences in the hydrodynamic solvers used to run the two simulations; due to differences in the modeling and implementation of CGM heating and cooling \citep[cf.~\S 2.1 of][for additional details]{jung2022massive}; or some combination thereof.  Whatever the reason, it also results in {\sc Romulus} and TNG CGM group-scale gas entropy profiles having very different shapes (see \citealt{oppenheimer2021simulating} for a detailed discussion); the shapes of the {\sc Romulus} entropy profiles are in good agreement with the observations \citep{jung2022massive}.

We also compare our results to \citet{esmerian2021thermal}. The profiles in their Fig. 3 are not directly comparable to those in our Fig. \ref{clogs} (left panel) because their halos are lower mass, Milky Way-like systems and their definition of the hot intragroup gas is based on an entropy cut ( K > 5 $\rm keV cm^2$) as opposed to our temperature cut ( T > $10^6 \rm K$).   However, when we apply the  Esmerian {\it et al.} entropy threshold and repeat the analysis, the resulting $t_\mathrm{cool} / t_\mathrm{ff}$ profiles, especially those of our lower mass systems, are in good agreement.   Our collective results sit on the same continuum.

\subsubsection{The condensation-susceptible CGM}\label{tuCGM}

Even a quick perusal of Fig. \ref{Fig2} shows that gas that is susceptible to condensation is \emph{not restricted only to regions where the median curve falls below} $t_\mathrm{cool}/t_\mathrm{ff} = 20$.  This is readily apparent in, for example, the {\sc Romulus C} panel.  Additionally, there are also radial zones, in halos 99966 and 38182 for example, where the median curve is below our threshold and yet, there is very little gas condensing out.   Features like this in the observations have been use to argue that the $t_\mathrm{cool}$ may be a better indicator of thermal instability than  $t_\mathrm{cool}/t_\mathrm{ff}$ \citep{hogan2017onset}.  Specifically, \citet{hogan2017onset} argue that the nebular emission, a tracer of cold gas and star formation, mainly occurs in systems with median $t_\mathrm{cool}$ < 1 Gyr at 10 kpc.  We find that nearly all {\sc Romulus} groups, both CCs and NCCs, have $t_\mathrm{cool}$ < 1 Gyr at 10 kpc.  However, not all halos' BGGs have ongoing star formation within the central 50 kpc sphere 
(see Table \ref{table:tuCGMMCGM}).  Moreover, the recently published observed profiles of \citet{martz2020thermally} are consistent with the {\sc Romulus} results.  They find a number of groups and clusters that have $t_\mathrm{cool}$ < 1 Gyr in the central region but do not evidence star formation or molecular gas in and about their central galaxy.  In light of this, we choose to stick with $t_\mathrm{cool}/t_\mathrm{ff}$.

Nonetheless, the existence of gas that is susceptible to condensation in radial zones where the median curve is above our threshold begs an explanation. We interpret such features as supporting the results of \citet{choudhury2019multiphase}, who investigated conditions under which both large and small density perturbations can become thermally unstable.  They found that the crucial parameter signaling the medium's susceptibility to multiphase condensations is not whether the median  $t_\mathrm{cool} / t_\mathrm{ff}$ is below some value but whether the perturbations' local $t_\mathrm{cool} / t_\mathrm{ff}$ falls below threshold, particularly if the perturbations are strong.
And, as we will show in \S \ref{origin} and \ref{thermodynamics}, gas in groups embedded in cosmologically realistic environments are regularly subject to strong perturbations. 

As for the lack of condensing gas when the median curve is below our threshold, this can be due to lack of perturbations.  However, a more likely explanation is that the picture sketched out here needs to be updated to account for transient heating episodes.  During such episodes, even if a perturbation has short cooling time, it will not condense out if its radiative losses are offset by heating.  A snapshot of $t_\mathrm{cool}$ or $t_\mathrm{cool}/t_\mathrm{ff}$ at a single moment will not reveal whether this is happening.  However, as shown in Figs. 13 -- 18 of \citet{jung2022massive}, all {\sc Romulus} groups experience repeated episodic AGN outbursts.

Guided by these findings, we focus our analysis on \emph{local} $t_\mathrm{cool} / t_\mathrm{ff}$ in CGM. In Fig.~\ref{Fig2}, we have shaded the CGM gas with local $t_\mathrm{cool} / t_\mathrm{ff}$ below 20 at the redshift of analysis, $z_{\rm a}$, translucent orange and will hereafter refer to this gas as ``condensation-susceptible'' CGM. The expectation is that the gas elements in this regime are susceptible to rapid cooling, or are actually in the process of doing so.

Note that although we are mainly interested in gas within the radial range $0.1 \leq \rm R/R_\mathrm{500} \leq 1$, the translucent orange component in the panels extends inward to $0.02 \rm R_\mathrm{500}$. This is because we also want to capture cooling CGM gas that has fallen or flowed in past inner boundary
but has not yet been incorporated into the BGG's ISM at $z_{\rm a}$.  Specifically, we designate gas 
in the radial range $0.02 \leq \rm R/R_\mathrm{500} < 0.1$ as ``condensation-susceptible'' CGM if it satisfies three conditions: (i) its $t_\mathrm{cool}/t_\mathrm{ff} \leq 20$ at $z_{\rm a}$; (ii) its temperature is greater than $ 5\times 10^4$ K  and its hydrogen number density is less than $0.1\;\mathrm{cm}^{-3}$ (i.e.~it is not part of the BGG's ISM); and (iii) it was part of the CGM (i.e.,~in the region $0.1 \leq \rm R/R_\mathrm{500} \leq 1$ at some point in the 5 Gyrs prior to the time of analysis).


\begin{table}
\fontsize{9}{11}\selectfont
\centering
\begin{tabular}
{|l|p{0.9cm}|p{0.5cm}|p{0.9cm}|p{0.9cm}|p{0.9cm}|p{0.9cm}}
 \hline
 \hline
 \\
 Halo ID & CC/NCC status & $z_\mathrm{a}$ & SFR [M$_\mathrm{\odot}$/yr] & $ \rm M_\mathrm{500}$ [M$_\mathrm{\odot}$] & M$_\mathrm{CGM}$ [M$_\mathrm{\odot}$] &  $f_{\substack{\rm condensation\\ \rm susceptible\\ \rm CGM}}$
\\
 \hline
 \\
 
C  & CC & 0.7 & 142.8 &4.6e+13 & 5.3e+12 &13.2\%  \\
52024  &CC & 0.29 &73.9 &1.0e+13 & 1.1e+12  &32.0\%  \\
G1  & CC &0.25 &21.3 &1.1e+13 & 1.2e+12  &37.8\%   \\
49510  &NCC & 0.36 &0.0 &8.2e+12 & 8.7e+11  &27.9\%   \\
99966  &CC & 0.31 &22.3 &3.9e+12 & 4.1e+11  &66.7\%  \\
38182  &NCC & 0.44 &0.0 &2.8e+12 & 2.6e+11  &74.0\% \\
91655  &NCC &  0.26 &0.64 &1.7e+12 & 1.1e+11  &66.9\% \\
77876  &NCC & 0.25 &8.6 &1.8e+12 & 1.5e+11  &88.0\% \\

 \hline
\end{tabular}
\
\caption{Properties of the eight {\sc Romulus} halos used in this study at $z_\mathrm{a}$. CC/NCC status identifies the system as a cool core or non-cool group based on the classification criterion described in \S \ref{BGG}. SFR is the BGG's star formation rate at $z_\mathrm{a}$ measured within a 50 kpc sphere around the halo center.
M$_\mathrm{500}$ is the total mass at $z_\mathrm{a}$ within R$_\mathrm{500}$. M$_\mathrm{CGM}$ is total gas mass of CGM. $f_\mathrm{condensation-susceptible~ CGM}$ = $\rm M_\mathrm{condensation-susceptible ~ CGM}/M_\mathrm{CGM}$ is the fraction of CGM gas mass that meets the ``condensation-susceptible'' CGM criteria.}
\label{table:tuCGMMCGM}
\end{table}


   \begin{figure*}
   \centering
   \includegraphics[width= 0.8\linewidth] {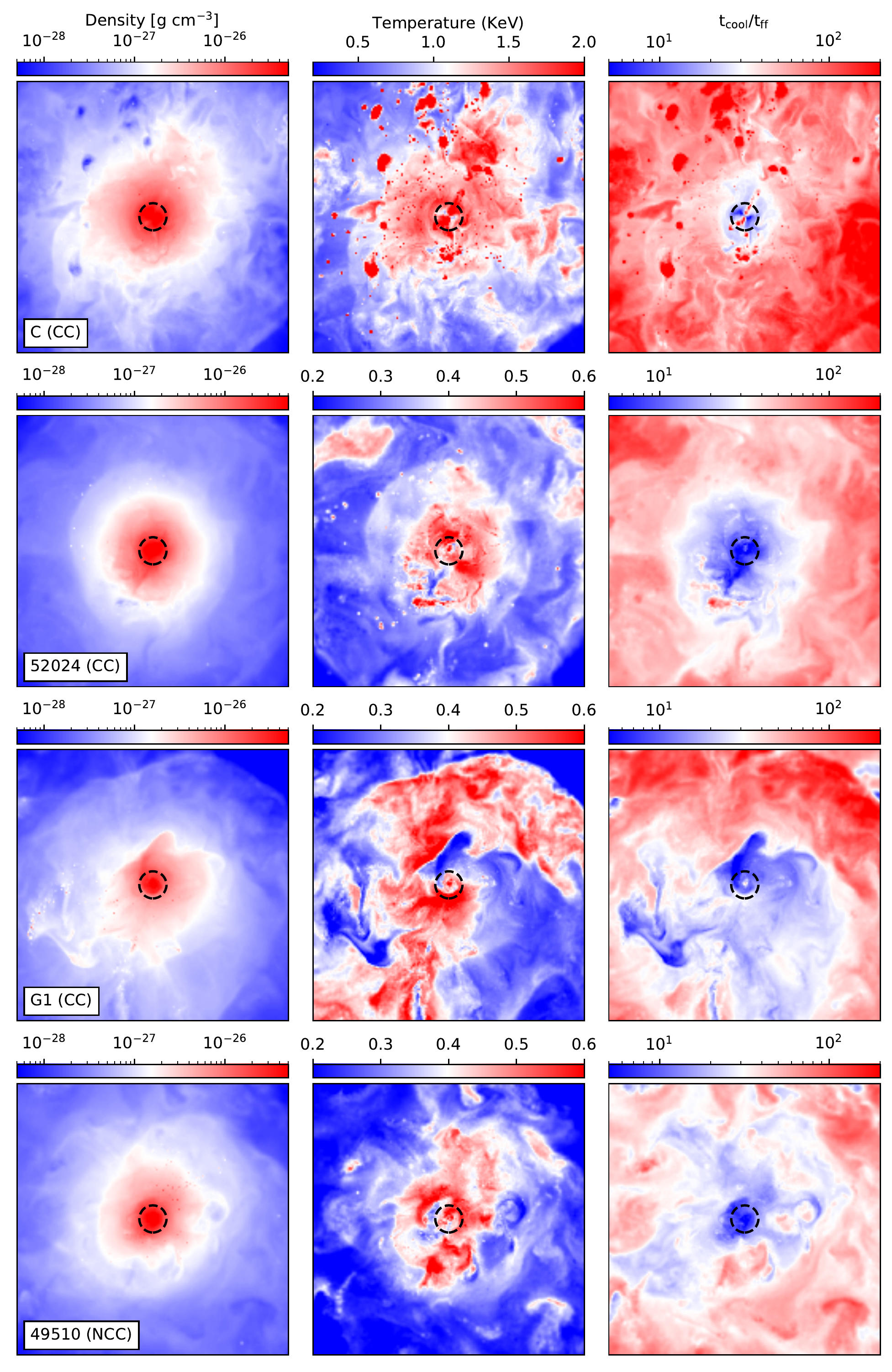}
      \caption{ Each row shows maps of projected density, temperature and $t_\mathrm{cool} / t_\mathrm{ff}$ of the CGM in a 30 kpc slice through the halo center for the four halos in Fig.\ref{Fig2}, {\sc Romulus} C (CC), 52024 (CC), G1 (CC), and 49510 (NCC) from top to bottom, at $z_\mathrm{a}$.Panels are $\rm 2R_\mathrm{500}$ on each side. The dashed black circle shows r = 0.1R$_{500}$. We only consider the gas beyond this distance as CGM. The other four halos are shown in Fig. \ref{projections02}.  }
         \label{projections01}
   \end{figure*}

   \begin{figure*}
   \centering
   \includegraphics[width= 0.8\linewidth] {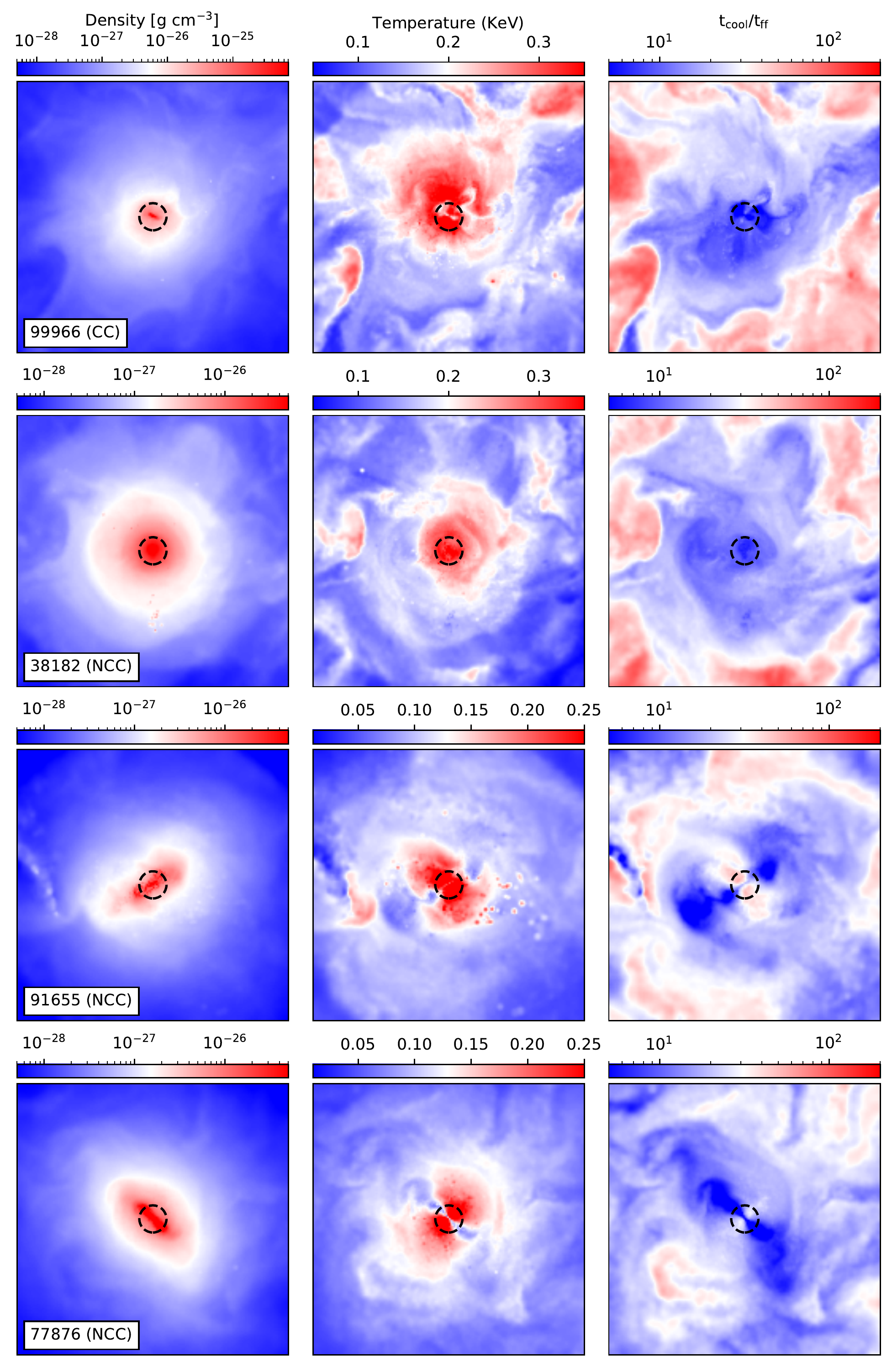}
      \caption{ Same as Fig. \ref{projections01} for halos 99966 (CC),38182 (NCC), 91655 (NCC), and 77876 (NCC) from top to bottom, at $z_\mathrm{a}$.
      }
         \label{projections02}
   \end{figure*}

Next, we mentioned above that some of the {\sc Romulus} BGGs are star forming while others are not.  We list the BGGs' star formation rates in Table \ref{table:tuCGMMCGM}, along with some of the other group/BGG properties including the groups' CC/NCC status. Of the eight halos we are studying in detail, four are CC and four are NCC groups.
We note a strong correlation between the central group galaxy's star formation rate (SFR) and its CC/NCC status.

To understand this correlation, we focus first on $\rm R > 0.1 \rm R_{500}$ gas in the groups.
We find that the thermodynamic properties of the gas in both CC and NCC groups in this region are similar. This is demonstrated by the similarity of the median profiles for various quantities, such as  $ t_\mathrm{cool} / t_\mathrm{ff} $ (right panel of Fig. \ref{clogs}) as well as the scaled entropy (left panel of Fig 12 of \citealt{jung2022massive}), in this region. These median profiles increase power-law-like towards larger radii, with nearly similar slopes in all halos, regardless of their CC/NCC status. We also looked at the fraction of the CGM that is identified as ``condensation-susceptible'' (last column in Table \ref{table:tuCGMMCGM}) and find that it too is not correlated with the CC/NCC status of the groups.

The main differences between CC and NCC groups are in the region $\rm R < 0.1 \rm R_{500}$: CC groups have a greater amount of gas with $t_\mathrm{cool}/t_\mathrm{ff}$ < 20 that has flowed into this region compared to NCC groups.  This can be seen in the high contrast and the spread of the grey shading as well as the amount of orange-shaded gas in CC and NCC groups at $\rm R < 0.1 R_\mathrm{500}$ in Fig.\ref{Fig2}. 
This deficit shows that  the CGM gas in NCC groups is not cooling onto the BGGs, which would explain the BGGs' low/negligible star formation rates. In detail, the cooling and/or inflowing gas in NCC groups is heated by AGN feedback as it gets closer to the BGG.  The impact of this heating is reflected, as already noted, in the groups' temperature profiles:  CC groups' temperature profiles manifest a drop towards the centre  while the NCC groups' temperature profiles are either flat or centrally peaked.  Also, the median $t_\mathrm{cool} / t_\mathrm{ff}$ profiles of the CC groups are slightly steeper in the central region than those of NCC groups.  For further discussion of CC and NCC groups see \S \ref{preexisting-CCNCC}.

\section{The nature and the origins of the CGM}\label{origin}

To gain further insights about the CGM in our group halos, we highlight in Figs. \ref{projections01} $\&$ \ref{projections02} the spatial structure in the CGM of the systems shown in Fig. \ref{Fig2}. From left to right, the columns show the distribution of mass-weighted average (along the line of sight) gas density, temperature, and $t_\mathrm{cool} / t_\mathrm{ff}$ ratio in slices with cross-sectional area, $\rm 2R_{\rm 500} \times 2R_{\rm 500}$, and $30$ kpc thick, centered on the BGG.  The black circle in each panel marks the central region of radius 0.1$\rm R_{\rm 500}$, the region we exclude when defining the CGM.

The inhomogeneous nature of the CGM in these halos is not obvious in the density panels due to projection effects. Mostly, the trends seen in the density panels are fairly typical of the structure present in all  {\sc Romulus} groups: gas density is the highest at center and drops off with distance from the center.  In all but the last row, the gas distribution is generally circularly symmetric.   Nonetheless, there are indications of churning due to mergers and AGN outflows.  For example, the ellipsoidal distribution in halos 91655 and 77876 in Fig. \ref{projections02} is the result of ongoing AGN jets and {\sc Romulus C} slice (first row of Fig. \ref{projections01}) contains numerous hot, low-density bubbles that are remnants of recent episodes of AGN feedback.

The inhomogeneities due to filamentary extensions as well as those due to AGN jets and jet-heated cavities show up prominently in the temperature (middle column) and $t_\mathrm{cool} / t_\mathrm{ff}$ (last column) panels.  This includes the filament of  cool gas extending to $\rm R_\mathrm{500}$ in the 
first two rows of Fig. \ref{projections02}; the numerous hot, low-density AGN-inflated bubbles in the first row of Fig. \ref{projections01}; and the wide bipolar AGN outflow cones in the last two rows of Fig. \ref{projections02}\footnote{See also Figs. 10 and 11 of \citealt{tremmel2017romulus} and Figs. 2 and 3 of \citealt{chadayammuri2020fountains} for different views highlighting the various accretion- and AGN-induced structures in {\sc Romulus C}.}. We note that although half of the systems are labeled as cool core in Table \ref{table:table1}, the declining gas temperature towards the group center is not evident in the panels because of the superposition of hotter gas along the line of sight.

The  $t_\mathrm{cool} / t_\mathrm{ff}$ ratio in Figs. \ref{projections01} $\&$ \ref{projections02} ranges from a few to 200, with the blue colour corresponding to ``condensation-susceptible'' gas.  This ``condensation-susceptible'' gas is ubiquitous within R$_\mathrm{500}$ of all halos in Romulus simulations and its distribution is amorphous and filamentary.  
Some of these filaments are due to cool inflowing gas in the plane perpendicular to the AGN outflows cones.  See, for example, panels for halos 91655 and 77876, in Fig. \ref{projections02}. Such features highlight that hot outflows and cool inflows can co-exist in the same region.\footnote{These features also highlight the difficulty of heating the CGM isotropically via narrow bipolar outflows, leading to suggestions that a more effective implementation for AGN jets is to ensure that they change direction every so often \citep{Babul2013jets,Cielo2018jets}.  This is in fact how the jets behave in the {\sc Romulus} simulations \citep[cf. Fig. 11 of][]{tremmel2019introducing}.}  As noted previously, we also see streams of unstable gas extending to R$_\mathrm{500}$ (see, for example, the first and second row of Fig. \ref{projections02}).  These are either inflowing filaments of cold gas penetrating deep into the halos --- such features are also present in the CGM of lower mass, Milky Way-like sytems as well (cf Figs. 11 and 13 of \citealt{sokolowska2018complementary}) --- or tails of cool gas stripped from infalling or orbiting satellites.  The latter are similar to features in Figs. 2-9 of \citet{poole2006mergingI}, who show that the tail of stripped gas initially has very different thermodynamic properties than the background medium.

We will discuss these various structures of ``condensation-susceptible'' gas in the group-scale halos in greater detail in section \S \ref{thermodynamics}.  We expect that some of this gas is condensing and contributing to the multiphase structure of the CGM. Additionally, 
orbiting and infalling substructures within halos engender wakes (see \citealt{kim2007heating,ghazvini2008stirring,Ruszkowski2011stirring,
tamfall2021gravwakes}, and references therein) and again, it would not be a surprise if these induce strong perturbations in the background CGM, driving some of that gas to condense as well.

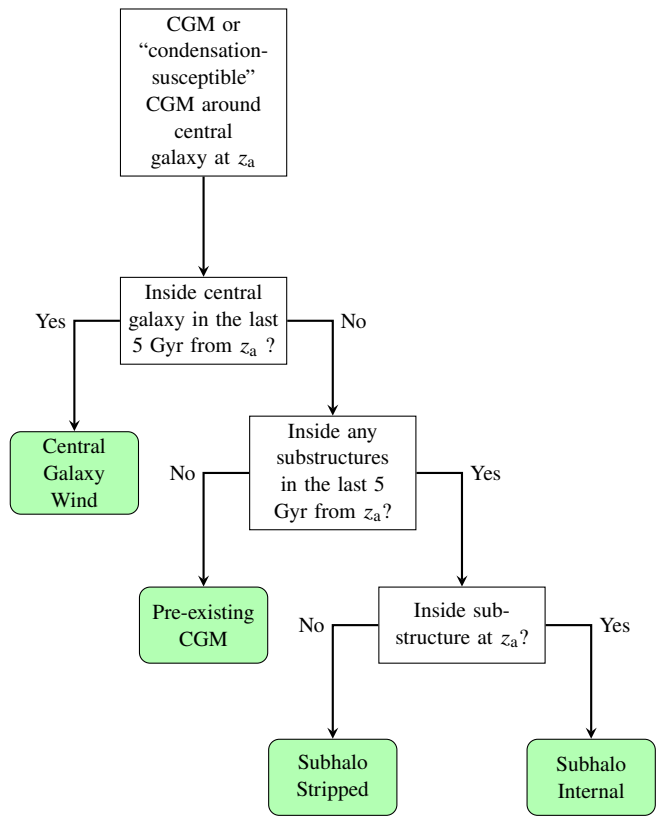
\begin{figure}
    \centering
    \begin{tikzpicture}[node distance=2cm]
    
    \node (start) [All] {CGM or ``condensation-susceptible'' CGM around central galaxy at $z_\mathrm{a}$};
    \node (q1) [All, below of = start, yshift = -1cm] {Inside central galaxy in the last 5 Gyr from $z_\mathrm{a}$ ?};
    \node (wind) [process,below of = q1, xshift = -1.7cm] {Central Galaxy Wind};
    \node (q2) [All,below of = q1, xshift = 1.7cm] {Inside any substructures in the last 5 Gyr from $z_\mathrm{a}$?};
    \node (q3) [All,below of = q2, xshift = 1.7cm] {Inside substructure at $z_\mathrm{a}$?};
    \node (cgm) [process,below of = q2, xshift = -1.7cm] {Pre-existing CGM};
    \node (present subhalo) [process,below of = q3, xshift = 1.7cm] {Subhalo Internal};
    \node (previous subhalo) [process,below of = q3, xshift = -1.7cm] {Subhalo Stripped};

    \draw [arrow] (start) -- (q1);
    \draw [arrow] (q1) -| node[left]{Yes} (wind);
    \draw [arrow] (q1) -| node[right]{No} (q2);
    \draw [arrow] (q2) -| node[right]{Yes} (q3);
    \draw [arrow] (q2) -| node[left]{No} (cgm);
    \draw [arrow] (q3) -| node[right]{Yes} (present subhalo);
    \draw [arrow] (q3) -| node[left]{No} (previous subhalo);
    \end{tikzpicture}
    \caption{Flow chart summarising how we classify the CGM and the ``condensation-susceptible'' CGM (orange gas in Fig. \ref{Fig2}). These are classified into 4 categories: \emph{Central galaxy wind}, \emph{subhalo internal}, \emph{subhalo stripped}, and \emph{pre-existing CGM}, based on its past history.  See the text for a more detailed definition of each category.
    }
    \label{flowchart}

\end{figure}
  \begin{figure}
   \centering
   \includegraphics[width=1.02\linewidth]{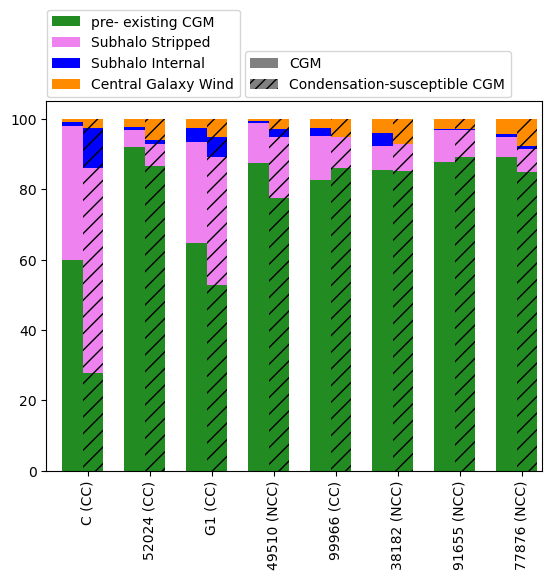}
      \caption{ The CGM and the ``condensation-susceptible'' CGM at $z_\mathrm{a}$ are grouped into 4 categories based on their history over the preceding 5 Gyrs, as described in the text. This plot shows the fraction of CGM (solid bars) and the ``condensation-susceptible'' CGM (hatched bars) corresponding to the four categories.  
      }
         \label{barchart34}
  \end{figure}


\subsection{CGM tracking and classification}\label{4classes}
In the previous section, we inferred that the CGM gas has likely come from multiple sources. 
In this section, we investigate the origin of the CGM as well as the  ``condensation-susceptible'' CGM at $z_{\rm a}$ by tracking the associated gas elements back in time.   We consider four potential sources, which are as follows:

\begin{itemize}[leftmargin=0.2cm]
\item \emph{Central Galaxy Wind}: Gas in the region of interest that was part of the BGG's ISM (i.e.~its halocentric distance was $<\rm R_\mathrm{gal}$, its temperature was  T < $5\times 10^4$ K  and its hydrogen number density was $n_H > 0.1  \, \mathrm{cm}^{-3}$) at any point during the 5 Gyrs immediately preceding $z_\mathrm{a}$.   For an explanation for why we have adopted a 5 Gyrs timescale, we refer the reader to \S\ref{subhalosection}.
\item \emph{Subhalo Internal}: Gas in the region of interest that has been brought there by infalling subhalos and is identified by AHF as still bound to a subhalo at $z_\mathrm{a}$. 
\item \emph{Subhalo Stripped}: Gas in the region of interest that was bound to a subhalo at some point during the 5 Gyrs preceding $z_{\rm a}$ but is no longer bound to the subhalo at $z_{\rm a}$.  In this study, we do not concern ourselves with how the gas was removed from the subhalo.  Possible mechanisms include expulsion from the subhalo via galactic winds, ram pressure stripping, and tails formed from debris of tidal interactions. 
\item \emph{Pre-existing CGM}: Gas that was neither in the central galaxy nor bound to any subhalos in the 5 Gyrs preceding $z_{\rm a}$.  Gas in this category includes that which was always inside the sphere of radius 
$\rm R_{\rm 500}$($z_{\rm a}$) over the 5 Gyrs window under consideration, as well as that which has accreted onto the region of interest from larger radii.
\end{itemize}
The flowchart in Fig. \ref{flowchart} summarizes our procedure for classifying the CGM and the ``condensation-susceptible'' CGM.  

In Fig. \ref{barchart34} we use bar charts to show the fractions of the CGM and the ``condensation-susceptible'' CGM contributed by the four sources, for our eight {\sc Romulus} groups.  Each individual bar chart has a hatched and an unhatched side showing the results for the CGM and the ``condensation-susceptible'' CGM, respectively.
 The fractions vary from halo to halo but broadly, the halos can be grouped into two categories: those with quiescent merger histories over the 5 Gyrs period preceding the time of analysis, and those with active merger histories, including possibly major mergers prior to the start of the window.   {\sc Romulus} C and G1 halos belong to the latter category.  

 For both sets of halos, the dominant component of the CGM is \emph{pre-existing CGM}.  In the quiescent halos, it makes up $>80\%$ of the CGM while in the active halos, the fraction is lower to $\sim 60\%$ while the \emph{subhalo stripped} has grown to $30-40\%$ vs. $\lsim 10\%$.  Neither the \emph{subhalo internal} nor the \emph{central galaxy wind} fractions make up more than $4\%$ generally.   The low fraction of the latter component may seem surprising but we note that as per our definition of the CGM, we do not consider gas at R < 0.1R$_\mathrm{500}$.

The low fraction of the \emph{central galaxy wind} material in the CGM may seem at odds with the results of \citet{hafen2019origins}. They investigated the origins of the CGM in halos with smaller galaxies (M$_\ast \simeq 10^{6} - 10^{11}\;$M$_\mathrm{\odot}$) than ours and found that typically,  the \emph{central galaxy wind} is the second most important component.  However, they also note that in their low redshift sample, the \emph{central galaxy wind} fraction decreases with increasing system mass, touching $10\%$ in their highest mass systems.  At the same time, their equivalent of our \emph{subhalo stripped} grows to $\sim 30\%$.  Our results are consistent with the extension of these trends to larger mass galaxies that we study.

Next, we consider the ``condensation-susceptible'' CGM.  In the active halos, the two dominant fractions are still those associated with the \emph{pre-existing CGM} and \emph{subhalo stripped} components but which of the two dominates varies.  In {\sc Romulus} C, \emph{subhalo stripped}  makes up the largest fraction while in {\sc Romulus} G1, \emph{pre-existing CGM} dominates. In the quiescent systems, \emph{pre-existing CGM} is still the dominant component and the corresponding fraction is similar to that in the CGM.  The fraction of the second largest category, \emph{subhalo stripped}, varies from being almost the same as in the CGM to slightly larger.  In both sets of halos,  the combined fraction of \emph{subhalo internal} and \emph{central galaxy wind} is generally larger for the ``condensation-susceptible'' CGM than the CGM, growing to as much $15\%$; however, as to which of the two is larger, there is no clear trend.  In some halos, \emph{central galaxy wind} is larger and in other halos, \emph{subhalo internal} is larger.

\section{Evolution of the condensation-susceptible CGM}\label{thermodynamics}

To better understand the origin and the fate of the ``condensation-susceptible'' CGM at the time of analysis, we carry out a detailed particle tracking analysis. Specifically, we tracked the gas two Gyrs back and two Gyrs forward in time from $z_a$, allowing us to analyze its behavior over a four-Gyr period. Based on analyses in this time span, we find that the gas can be grouped into seven sub-components whose evolution is discernibly different.

\begin{figure*}
   \centering
   \includegraphics[width=0.85\linewidth]{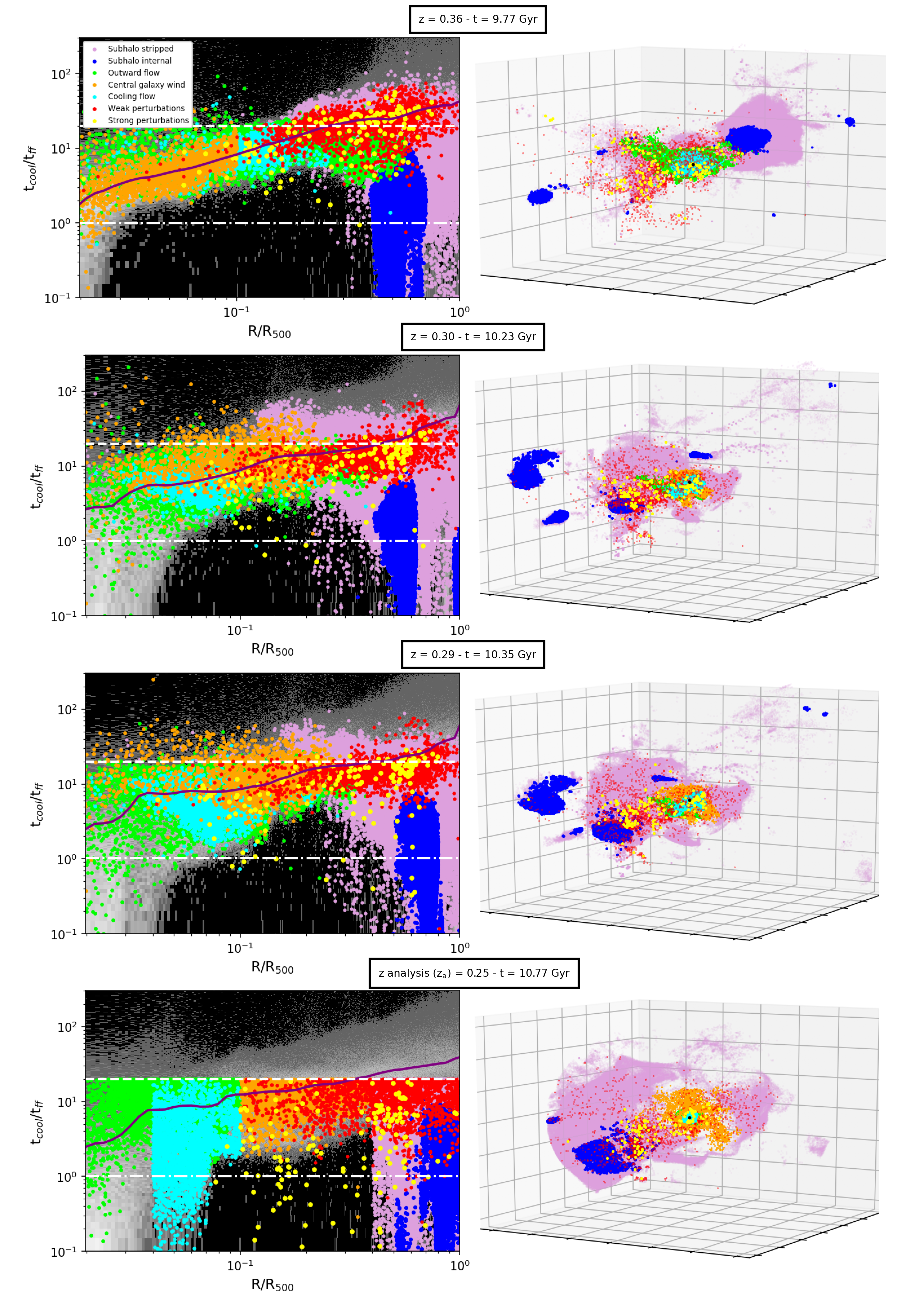}
      \caption{The ``condensation-susceptible'' at $z_\mathrm{a}$ (last row) comprises seven sub-components whose evolution is discernibly different (see text for details). Here, we use {\sc Romulus} G1 group to illustrate how these sub-components evolve. At $z_\mathrm{a}$, we sub-sample gas associated with each sub-component; plot them in different colors (see legend) and follow this gas backward and forward in time. The first three rows show the state of the sub-components at three earlier epochs.   
      The left column show the gas in the $t_\mathrm{cool} / t_\mathrm{ff}$ vs.~radius plane; the right columns show its 3D spatial distribution.
      }
         \label{tcooltff3D-RG1-p1}
  \end{figure*}

\begin{figure*}
   \centering
   \includegraphics[width=0.87\linewidth]{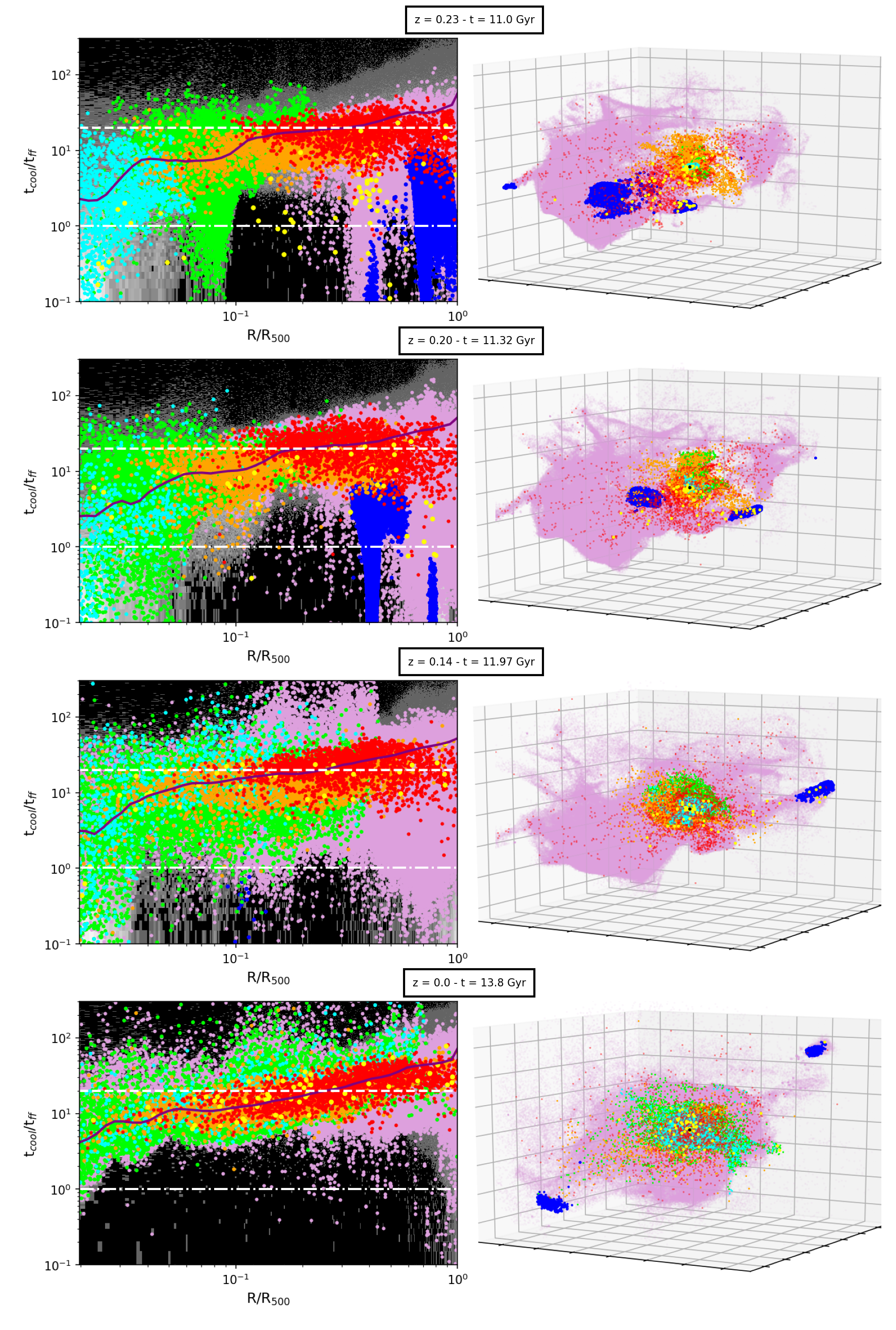}
      \caption{ This figure is the same as Fig. \ref{tcooltff3D-RG1-p2} except that the plots correspond to four epochs following $z_\mathrm{a}$.  See caption for Figure \ref{tcooltff3D-RG1-p1} and the text for details.
              }
         \label{tcooltff3D-RG1-p2}
  \end{figure*}

In Figs. \ref{tcooltff3D-RG1-p1} and \ref{tcooltff3D-RG1-p2}, we use {\sc Romulus} G1 halo as an example to highlight these seven sub-components and illustrate how they each evolve in the period leading up to and following the time of analysis.  Firstly, we direct attention to the last row in Fig. \ref{tcooltff3D-RG1-p1}.  This row corresponds to the time of analysis.  This is the epoch around which our analysis pivots. In the left panel, we use the seven different colors to highlight a small subset of the gas belonging to each of the sub-components in the $t_\mathrm{cool} / t_\mathrm{ff}$ vs. radius plane.  The main purpose of Figs. \ref{tcooltff3D-RG1-p1} and \ref{tcooltff3D-RG1-p2} is to clearly show the evolutionary trends.  It is for this reason that we  do not color all the gas because that results in a confusing plot with a hodgepodge of overlapping multi-coloured points.   To further minimize the overlapping and enhance clarity, each subsamples were also extracted from targeted radial ranges.  To that end, the distributions of the different colored points in the last row of Fig. \ref{tcooltff3D-RG1-p1} do not reflect either the mass fractions or the actual radial distributions of the corresponding sub-components.

\begin{itemize}[leftmargin=0.2cm]
\item The \arf{cyan} points sample the sub-component of the \emph{pre-existing CGM} that have a cooling flow-like behaviour.
\item The \arf{yellow} and \arf{red} points sample sub-components of the \emph{pre-existing CGM} that have been perturbed. We identify this gas by a sharp decrease in its $t_\mathrm{cool} / t_\mathrm{ff}$ within 0.5 Gyrs prior to $z_a$. We observe two distinct evolutionary patterns in this gas that correlate with the magnitude of the change in density. The yellow points correspond to gas with a significant increase in density, while the red points correspond to gas with a mild density change.
\item The \arf{lime-green} points sample gas that is moving outward at the time of analysis and has been doing so coherently for at least 0.5 Gyrs.
Since this gas has never been part of the central galaxy, we treat it separately from the \emph{central galaxy wind}.
\item The \arf{lavender} and \arf{blue} points are a subset of \emph{subhalo stripped} and \emph{subhalo internal} gas, respectively. 
\item The \arf{orange} points show a subset of the gas from the \emph{central galaxy wind} category. 
\end{itemize}
For completeness, we note that the grey shading and the purple curve shows, as before,  all the gas in the halos and the median $t_\mathrm{cool} / t_\mathrm{ff}$ profile for the hot gas.


Although we do not show all the gas belonging to these categories, we have assessed all of the ``condensation-susceptible'' CGM gas in all eight halos listed in Table \ref{table:table1} and found the same evolutionary patterns across all the halos.  

In the subsections below, we discuss the evolution of each of the sub-components in more detail but first, we briefly explain the organization of the rest of panels in Figs. \ref{tcooltff3D-RG1-p1} and \ref{tcooltff3D-RG1-p2}.  As noted, the bottom row in Fig. \ref{tcooltff3D-RG1-p1} corresponds to redshift of analysis, $z_\mathrm{a}$.   The preceding three rows show the distribution of the sampled gas at three epochs prior to $z_\mathrm{a}$  while the four rows in Fig. \ref{tcooltff3D-RG1-p2} show the results at four epochs post-$z_\mathrm{a}$.  The left panels show the changing gas distribution in the $t_\mathrm{cool} / t_\mathrm{ff}$ vs. radius plane and the right panels show the spatial distribution of the coloured points.

\subsection{{\arf{Orange}}: Central galaxy wind}

We first consider the \emph{central galaxy wind} sub-component of the ``condensation-susceptible'' CGM at $z_\mathrm{a}$.  A sample of gas from this sub-category is colored \arf{orange} in Figs.  \ref{tcooltff3D-RG1-p1} $\&$ \ref{tcooltff3D-RG1-p2}.  This is gas that was part of the BGG's ISM and then expelled from the BGG after being heated by the AGN at some point during the preceding five Gyrs. 
The first row in Fig. \ref{tcooltff3D-RG1-p1}  shows the state of this component roughly a gigayear before $z_\mathrm{a}$.  The gas is concentrated within the central core.  This is apparent in the left panel. In the right panel, the central concentration of \arf{orange} points is obscured by the overlapping points associated with the other components.     
As the subsequent panels show, this component expands outward.   The right panels show this gas breaking out of the central region. As it does so, it shocks, mixes, and about a gigayear after $z_\mathrm{a}$, essentially becomes part of the CGM.  Thereafter, it evolves like the CGM.  It is susceptible to perturbations and cooling but mostly it remains distributed about the median curve for hot CGM.

\subsection{{\arf{Blue}}: Subhalo internal and {\arf{Lavender}}: Subhalo stripped}\label{subhalosection}

Next, we consider the sub-components identified as \emph{subhalo internal} and \emph{subhalo stripped}.  The former is gas with $t_\mathrm{cool} / t_\mathrm{ff} < 20$ that, at $z_\mathrm{a}$, is bound to subhalos  while the latter is gas that was removed from the subhalos in the 5 Gyrs preceeding $z_\mathrm{a}$.  The \arf{blue} and \arf{lavender} points correspond to a subset of gas elements comprising these two categories. 

Typically, subhalos and their  gaseous tails (\arf{blue} and \arf{lavender} points) collectively enter the region of interest from the right in the left panels and move towards the center.  However, the process is not unidirectional.  Over the time period shown in Figs. \ref{tcooltff3D-RG1-p1} and \ref{tcooltff3D-RG1-p2}, many subhalos reach their pericenters and travel outward in radius. A small number of satellites, typically the more massive ones, merge with central galaxy and contribute their cold gas to the BGGs. This is one way the BGGs acquire fresh cold gas and rejuvenate (e.g., \citealt{olivares2022gas,Lagos2022,jung2022massive}). 

All orbiting subhalos eventually lose most of their gas.  Visualizing this evolution in the left panels is not straightforward, which is why we also plot the 3D spatial distribution of the gas in the right panels.  Consider, for example, the first three rows of Fig. \ref{tcooltff3D-RG1-p1}. One can see a number of satellites (\arf{blue} knots), make out their orbital paths, and observe gas loss in the form of \arf{blue} elongations transitioning to \arf{lavender} tails. The gas removed from the subhalos does not immediately detach from its source substructure nor does it immediate stop following the orbital trajectory it was on when stripped.  It takes time for the gas to slow down and detach; in the meantime, the gas manifests as streams trailing the satellites.

After detaching, some of the gas cools, loses angular momentum due to drag and eventually falls to the center. This is more likely to happen when the satellite's initial pericenter is close to the BGG.  And, this gas too contributes to the 
the re-emergence of gaseous and star forming disks/rings in the BGGs (e.g., \citealt{olivares2022gas,Lagos2022,jung2022massive}).
A larger fraction of the removed gas, however, shocks, heats up, mixes, and becomes part of the CGM.

\begin{figure}
   \centering
   \includegraphics[width=\linewidth]{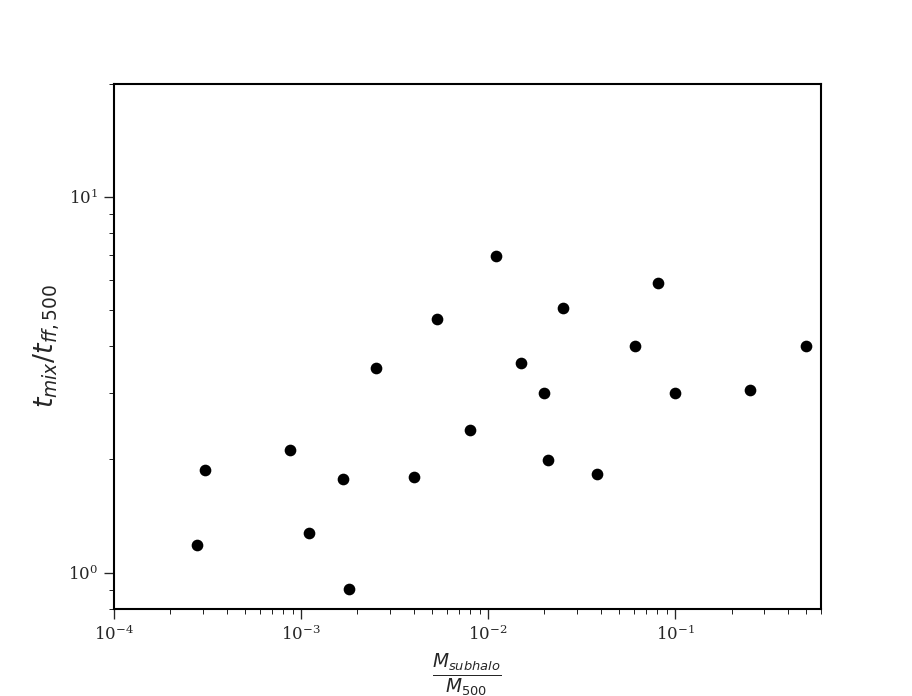}
      \caption{The amount of time that it takes for ``stripped'' gas from a substructure to mix with CGM. The y-axis shows mixing time over free-fall time at $\rm R_{500}$. The x-axis shows the substructure mass relative to its host halo's mass, $\rm M_{500}$. The scatter aside, the mixing time of the removed gas increases with mass fraction.
      }
         \label{Fig7}
   \end{figure}

We have determined the ``mixing time'' of gas from individual subhalos. The mixing time is defined as the time it takes for the entropy of 90\% of the gas removed from subhalos, and which is at R > 0.1R$_\mathrm{500}$, to reach an entropy level of more than 80\% of the median entropy at all radii. This time is measured from the moment when the subhalo crosses R$_\mathrm{500}$. In Fig. \ref{Fig7} we show this mixing time as a function of the free-fall time at R$_\mathrm{500}$ (i.e.,  t$_\mathrm{mix}$ / t$_\mathrm{ff,500}$) against the ratio of the subhalo mass to its host halo mass M$_\mathrm{500}$.  Both subhalo and halo mass are computed at the time subhalo is at R$_\mathrm{500}$.  To make this plot, we selected at least two subhalos in each halo (three, in the case of more massive halos) in a manner that reflects the range of subhalo-to-halo mass ratios in our simulations. We find that on the whole t$_\mathrm{mix} $/ t$_\mathrm{ff,500}$ increases with M$_{\mathrm{subhalo}}$/M$_\mathrm{500}$, albeit with large scatter. Nevertheless, most of the subhalo gas tends to mix within 5 Gyr. For this reason, when categorising the CGM we consider gas removed from a subhalo > 5 Gyr prior to $z_\mathrm{a}$ as \emph{pre-existing CGM}. For consistency we use the 5 Gyr look back window for all other categories as well.

   \begin{figure*}
   \centering
   \includegraphics[width=\linewidth]{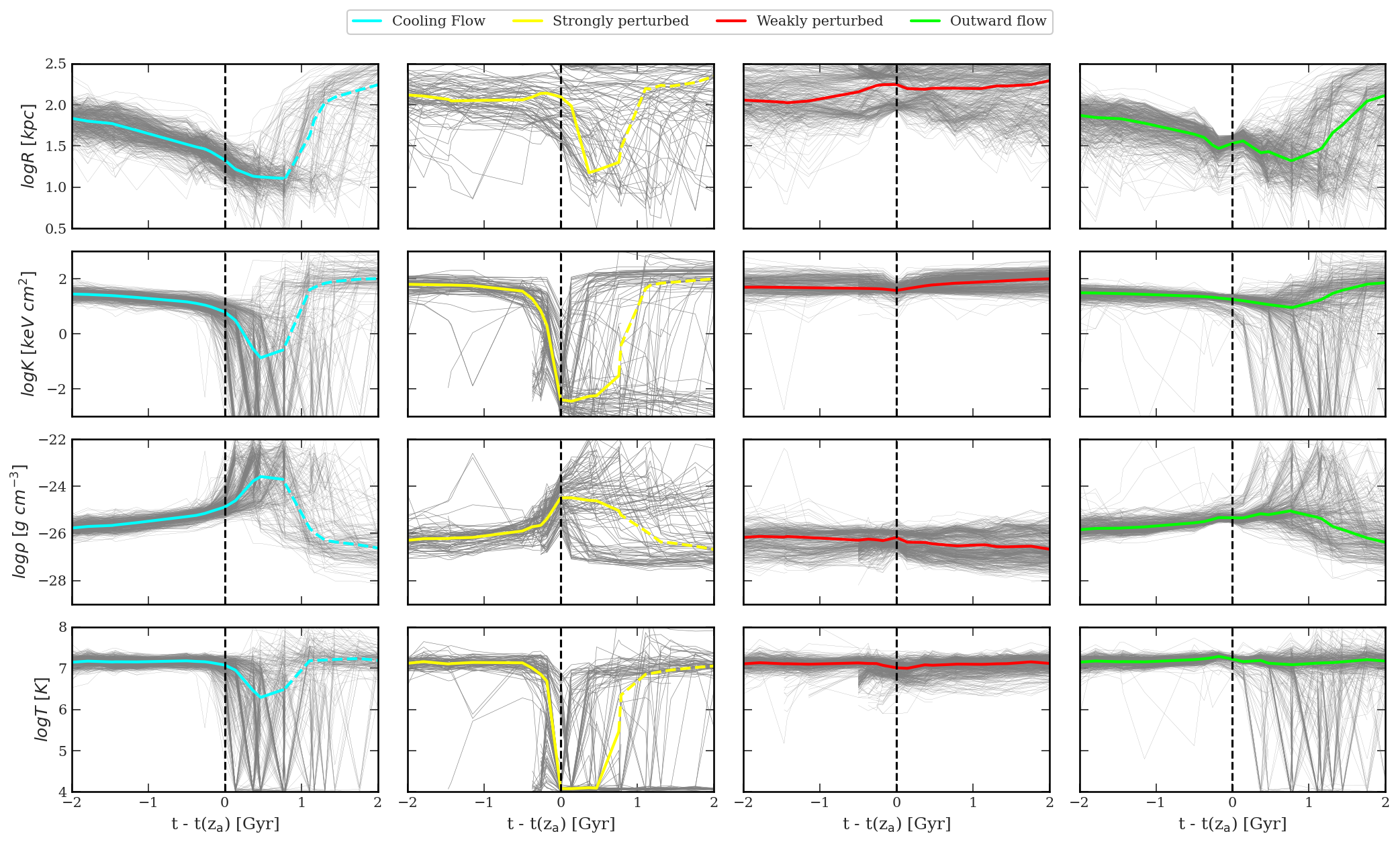}
      \caption{``condensation-susceptible'' gas in \emph{pre-existing CGM} category exhibits four different types of evolutionary behavior. This plot shows the thermodynamic history of the associated gas. Each column shows results for each evolutionary behavior. 
      \textbf{First column:} gas that demarcates the isothermal cooling flow. \textbf{Second column:} gas that has been subjected to strong perturbations. \textbf{Thirds column:} gas that has been subjected to weak perturbations. \textbf{Fourth column:} initially inward flowing CGM that got turned around and at the time of analysis is moving outward. The rows show radius, entropy, density, and temperature, from top to bottom. We follow the gas for two Gyrs before and after the analysis time t($z_\mathrm{a}$), shown by a black dashed line. Solid coloured lines show the median, and individual tracers are shown in the faint grey line. The dashed coloured line shows the median of the gas that is returned to the CGM via galactic outflow.}
         \label{thermohistory}
   \end{figure*}


\begin{table*}
\fontsize{9}{11}\selectfont
\centering
\begin{tabular}{lccccccc}
 \hline
 \hline
 \\
 Halo ID & CC/NCC status & $z_\mathrm{a}$ &  Cooling flow & Strongly perturbed & Weakly perturbed & Outward flow & Other \\
 \hline
 \\
 
C  & CC & 0.7  & 70.55\% & 1.3\% & 5.60\% &14.6\% &8.70\%\\
52024  & CC & 0.29 & 67.46\%&1.4\%&9.06\% &14.2\% &7.80\%\\
G1  & CC & 0.25 & 29.78\%&0.25\%&30.26\% &29.9\% &9.97\%\\
49510  & NCC & 0.36 & 33.80\%&0.27\%&25.88\% &32.32\% & 8.70\%\\
99966  & CC & 0.31 & 56.10\%&0.89\%&12.03\% &23.08\% & 7.9\%\\
38182  & NCC & 0.44 & 40.71\%&0.11\%&11.77\% &36.77\% & 10.7\%\\
91655  & NCC & 0.26 & 32.32\%&0.1\%&18.10\% &35.02\% & 14.5\%\\
77876  & NCC & 0.25 & 48.62\%&0.02\%&13.20\% &31.30\% & 7.1\%\
\\
 \hline
\end{tabular}
\
\caption{ The table shows the fraction of \emph{pre-existing CGM} that follows the four different evolutionary behaviours. In previous figures, the isothermal cooling flow, strongly and weakly perturbed gas, and outward flow are shown in \arf{cyan}, \arf{yellow} and \arf{red}, and \arf{lime-green}. Fractions of \emph{pre-existing CGM} that are not following any of these behaviours are shown in the last column.}
\label{table:3Bpre-existingCGM}
\end{table*}

\subsection{Pre-existing CGM}\label{tu-preexisting}
The final category is \emph{pre-existing CGM}. The gas in this category exhibits four broad evolutionary patterns. A subset of the gas from each of these four patterns are shown in Figs. \ref{tcooltff3D-RG1-p1} and \ref{tcooltff3D-RG1-p2} as \arf{cyan}, \arf{red}, \arf{yellow}, and \arf{lime-green} points.  In describing their behaviour, we will refer to Figs. \ref{tcooltff3D-RG1-p1} and \ref{tcooltff3D-RG1-p2} as well as Fig. \ref{thermohistory}.
Fig. \ref{thermohistory} shows the radius, entropy, density, and temperature of the sampled gas belonging to the four sub-components as a function of $\Delta$t $\equiv$ [t$-$t($z_\mathrm{a}$)].
The coloured lines show the median time evolution of the quantities, and that of the individual tracer trajectories are shown as thin grey lines in the background. Each sub-component is represented by the same colour in Fig. \ref{thermohistory} as in Figs. \ref{tcooltff3D-RG1-p1} and \ref{tcooltff3D-RG1-p2}.  And, while these plots are specific to {\sc Romulus} G1, we observe similar trends in all the halos. The fraction of the \emph{pre-existing CGM} in each of the four components, on a halo by halo basis, is given in Table \ref{table:3Bpre-existingCGM}. 

\subsubsection{\arf{Cyan:} Cooling flow} \label{coolingflow}
Firstly, we describe the evolution of the gas characterized by the \arf{cyan} points. 
The first row of Fig. \ref{tcooltff3D-RG1-p1} shows the state of the sampled gas $\sim$1 Gyr before $z_\mathrm{a}$.  The \arf{cyan} points are distributed around the median $t_\mathrm{cool} / t_\mathrm{ff}$ of hot CGM (purple line). 
Over time, this gas remains clustered about this median line as it flows inward.\footnote{The behaviour in the entropy-radius plot (not shown) is similar.  As the gas moves inwards, it remains clustered about the median entropy profile of the hot gas, which decreases towards the centre.}  This is clearly demonstrated by the cyan curve in the top left panel of Fig. \ref{thermohistory}. This is an example of an isothermal cooling flow \citep[cf~][]{Nulsen1998isothermalCF} in that  the temperature of this subcomponent does not change much leading up to $z_\mathrm{a}$ and even shortly thereafter (see bottom left panel in Fig. \ref{thermohistory}).   The entropy of the gas is, however, dropping (see the second panel in the first column of Fig. \ref{thermohistory}) and the flow is radiatively losing energy.  With respect to the temperature, compressional heating is largely balancing radiative losses.

During the above phase, the inward flow is subsonic, similar to that seen in the simulation discussed in \citet{lewis2000effects}. Also, as described in studies like those of \citet[][and references therein]{sokolowska2018complementary,stern2019cooling}, it is mainly localized to filamentary structures threading the inner halo (as apparent in the third row of Fig. \ref{projections01}) and therefore, does not involve all of the gas at any radius.  The fraction of gas in the cooling flow component in each of the eight halos is listed in Table \ref{table:3Bpre-existingCGM}.  The most apt description of the flow is a revised version of that by \citet{theuns_2003}: a slowly-moving filamentary emulsion of gaseous blobs with slightly different densities and temperatures [than the median] that is cooling isothermally and slipping past the other components comprising the CGM. 

Only after the gas enters the galaxy (i.e., R < R$_\mathrm{gal}$) does its time evolution bifurcate. Most of the gas starts to cool rapidly --- its density increases rapidly and both its entropy and temperature drop steeply --- and it becomes part of the BGG's ISM.  A small fraction, however, is halted and turned around by 
the outflowing galactic wind before rapid cooling happens (cf \S \ref{outward flow}).  The latter returns to the CGM.

The gas that settles in the central galaxy has non-zero angular momentum and is the dominant source of the gas leading to the re-emergence of the gaseous disks seen in \citet{jung2022massive}.  A fraction of this gas accretes onto the SMBH or fuels star formation, and a fraction is expelled as wind back into the CGM and eventually mixes with the ambient CGM.  The \arf{cyan} points in the last three panels of Fig. \ref{tcooltff3D-RG1-p2} correspond to both the pushed-back and the expelled gas; we preserve their colors to show that they were part of the cooling flow sub-component at $z_\mathrm{a}$.   And in Fig. \ref{thermohistory}, the 
median properties of \emph{this} gas is shown by dashed lines.  The dispersal of this gas to large radii is clearly shown in the top left panel.

In Table \ref{table:accreted-gas}, we show the fraction of the cooling flow that contributes to the star formation or accretes onto the SMBH.  Not surprisingly, the fraction is correlated with the halo's CC/NCC status; also, a larger fraction at $z_\mathrm{a}$ corresponds to a larger SFR.  We discuss these relationships further in  Section \ref{preexisting-CCNCC}.  

\subsubsection{\arf{Yellow} and \arf{Red}: Externally perturbed gas} \label{perturbedgas}

Next, we examine the sub-components of the \emph{pre-existing CGM} represented by the \arf{yellow} and \arf{red} points in Figs. \ref{tcooltff3D-RG1-p1} $\&$  \ref{tcooltff3D-RG1-p2}.  Both are examples of CGM gas that has been subjected to density perturbations. Various physical processes,
including shocks, gravitational focusing and turbulence due to orbiting substructures, as well as shocks and turbulence
caused by AGN- and SNe-powered galactic outflows, can (and do) induce density perturbations in the CGM. 
The forward evolution of these perturbations depend on whether locally they can cool efficiently, which in turn depends on their amplitudes \citep{choudhury2019multiphase,Das2021shatter}.\footnote{And in the case of simulations, also on whether their cooling lengthscale ($l_\mathrm{cool}=C_s t_\mathrm{cool}$) is resolved \citep{mandelker2021thermal}.}
The \arf{yellow} points sample the gas that experiences strong perturbations while the \arf{red} points sample the gas that is only mildly perturbed.  We characterize the perturbations as weak or strong on the basis of the increase in density (yellow and red curves in the third row of the second and third columns of Fig. \ref{thermohistory}) leading up to the time of analysis.  The median of the \arf{yellow} gas increases in density by a factor of $\sim$30 while the median of the \arf{red} gas registers only a slight increase, and only for a very short time.

\begin{figure}
   \centering
   \includegraphics[width=\linewidth]{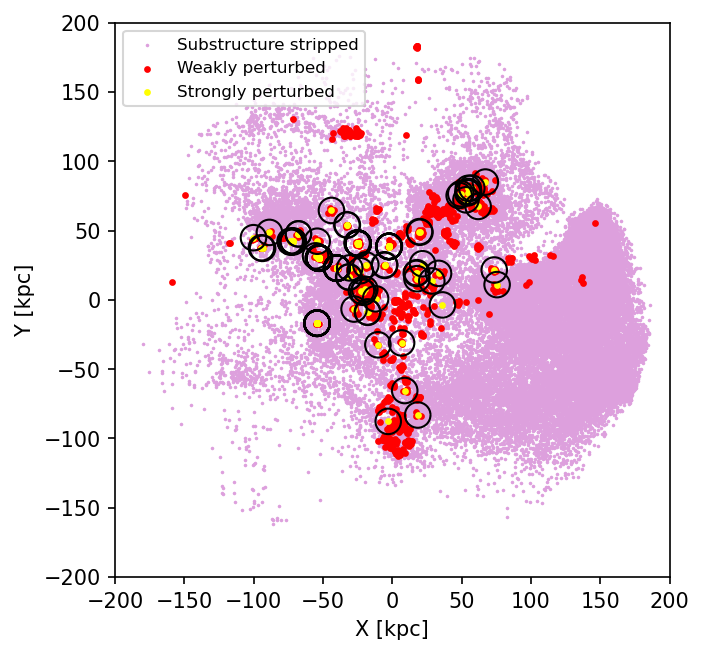}
      \caption{This 2D plot shows the position of the perturbed gas, shown as \arf{yellow} and \arf{red} points, and the subhalo stripped gas, shown in \arf{lavender}. The black circles denote the location of the neighbouring particles within the SPH kernel of each \arf{yellow} point. The plot demonstrates that strongly perturbed gas is consistently surrounded by gas that has been removed from subhalos and weakly perturbed CGM.
              }
         \label{subgroup-pos}
   \end{figure}

Before discussing the behaviour of the density perturbations in more detail, it is important to address a potential concern that the \arf{yellow} and \arf{red} particles are not the products of physical processes but of numerical noise, especially given the small fraction of mass involved (cf.~columns 5 and 6 of Table \ref{table:3Bpre-existingCGM}).  We check this in two ways: (i) Strictly speaking,  a single tracer SPH particle is a non-independent resolution element in that although the particles carry physical properties, such as mass, density, volume, velocity, temperature, etc., the physical information about the fluid and the flow is actually carried by a cloud of particles comprising the particle and its nearest neighbours \citep{agertz2007fundamental,sigalotti2021mathematics}.  We therefore compare the cooling time  of individual tracer particles to the average $t_\mathrm{cool}$ of a fluid element comprising all the particles within the tracer particle's SPH kernel.  We find that for 70\% of the tracer particles, their neighborhood's average $t_\mathrm{cool}$ is either statistically the same as the particle's cooling time or smaller (i.e.~the neighbourhood is collectively cooling similarly or faster).  This strongly suggests that majority of our yellow and red particles are not random, isolated points.  They are part of a region that itself is prone to condensation.  (ii) Additionally, we examine the spatial positions of the \arf{red} and \arf{yellow} particles in our halos.  Fig. \ref{subgroup-pos} shows that these tracers of perturbed \emph{pre-existing CGM} are mainly found in and around streams of gas from the subhalos (shown as \arf{lavendar} points), lending support to the idea that the main sources of the perturbations are the wakes and tails of orbiting subhalos.  This was already suggested by the distributions of the \arf{yellow} and \arf{red} points in the right column of Figs. \ref{tcooltff3D-RG1-p1} and \ref{tcooltff3D-RG1-p2}.

Focusing first on the strong perturbations, we note that in 
Figs. \ref{tcooltff3D-RG1-p1} $\&$  \ref{tcooltff3D-RG1-p2}, the \arf{yellow} points are sampling gas that starts at the median line.  At $\Delta t \approx -0.3$ Gyr, the gas is  
perturbed mainly by the satellites and experiences a strong density enhancement with respect to its local background.  In the second row of Fig. \ref{tcooltff3D-RG1-p1} a cluster of \arf{yellow} points forms close to a satellite (\arf{blue} points). This clustering can also be seen in the right panel (3D plots) overlapping the satellite and its gaseous tail.  These perturbations cool rapidly; the temperature of the gas drops --- from $ \sim 10^{7}$ to $\sim 10^{4}$ K --- and so does its entropy as well as its $t_\mathrm{cool} / t_\mathrm{ff}$ (see the panels in the second column of Fig. \ref{thermohistory}).  By $z_\mathrm{a}$, $t_\mathrm{cool} / t_\mathrm{ff}$ of all the perturbations is below 20 (dashed line), and some are even below 1 (dash-dotted line). Eventually, all of these perturbations end up below $t_\mathrm{cool} / t_\mathrm{ff} = 1$. 

Thereafter, some of the condensations fall in, some ride the satellites' wake for $\sim 1-2$ Gyrs before detaching and falling to the center, and some appear to be in stable orbits (see the first three 3D panels in Fig. \ref{tcooltff3D-RG1-p2} as well as the individual tracks in the first panel of the second column of Fig. \ref{thermohistory}).  

Most of the infalling gas joins the BGG's ISM.  We will refer to this cold infalling gas as ``precipitation''.  Once in the BGG, some of precipitation contributes to star formation and fueling the SMBHs, like the cooling flow gas, and some is expelled back to the CGM, as illustrated by the yellow points in the last two panel of Fig. \ref{tcooltff3D-RG1-p2}. 

A more detailed analysis shows that while the substructures are the main source of large-amplitude density perturbations in the CGM of galaxies in cosmological simulations,  the presence of the \arf{yellow} points in the same region as the \arf{orange} wind points in the left panel of third and fourth rows of Fig. \ref{tcooltff3D-RG1-p1} also suggests that outflows can perturb the gas, as seen previously in idealized simulations \citep[cf.~][]{prasad2015cool}.

Next, we consider the \arf{red} points. Like the strongly perturbed gas, this gas component is also initially distributed about the median purple line and it too is perturbed by the satellites.  This can be seen in the last 3D plot in Fig. \ref{tcooltff3D-RG1-p1} and the first and second 3D plots of Fig. \ref{tcooltff3D-RG1-p2}, where \arf{red} points are in the tail of \arf{blue} substructures.   However, unlike the strongly perturbed gas, most of this gas is swept up and carried along in the satellite wakes. Additionally, while the $t_\mathrm{cool} / t_\mathrm{ff}$ of these perturbations initially drop and even fall below our threshold of 20 by $z_\mathrm{a}$, unlike the strongly perturbed gas, they do not enter a runaway cooling state.  Instead, their ratios eventually bounce and return to the median value for the hot gas.  This bounce is visible in going from the last row of Fig.  \ref{tcooltff3D-RG1-p1} to the first row of Fig. \ref{tcooltff3D-RG1-p2}.

Given the findings of \citet{mandelker2021thermal}, one cannot help but wonder whether these perturbations are physically stable or whether they are stable due to numerical reasons?  \citet{mandelker2021thermal} argue that under certain conditions, small-amplitude perturbations will not be able to cool and condense if their cooling lengths ($l_\mathrm{cool} = \rm c_s t_\mathrm{cool}$) are not resolved.  We tested whether this was the case and found that most of the weak perturbations' cooling lengths are in fact resolved.

Another possibility is that only perturbations whose local $t_\mathrm{cool} / t_\mathrm{ff}$ ratios drop below unity truly cool and condense \citep{voit2021graphical}.   We have tracked the evolution of the weak perturbations and find that this does indeed appear to be the case.  In fact, $\gsim$ 95\% of the perturbations bounce before reaching $t_\mathrm{cool} / t_\mathrm{ff} = 2$.  And, while we find that $\sim$1\% of perturbations drop below unity before bouncing, the $l_\mathrm{cool}$ of nearly all of these is not resolved.

To understand the ``bouncing'' perturbations better, we examine their properties more closely.
The first panel of the third column of Fig. \ref{thermohistory} shows that the gas is first swept up in satellite wakes roughly a Gyr prior to t($z_\mathrm{a}$).   From this point on, the overall density of the gas (third row, third column of Fig. \ref{thermohistory}) decreases steadily but gently, except for a brief period of compression at t($z_\mathrm{a}$).  In the second panel, we note that the gas entropy leading up to t($z_\mathrm{a}$) increases, albeit slightly.  At t($z_\mathrm{a}$), we witness a slight dip, and then it continues to rise.  Cooling time ($t_\mathrm{cool}$) too behaves similarly.  Simply put, we do not see the perturbations' cooling times shortening. Then why does  $t_\mathrm{cool} / t_\mathrm{ff}$ drop?

Most of the studies investigating $t_\mathrm{cool} / t_\mathrm{ff}$ of gas perturbations assume that the ratio varies due to decreasing $t_\mathrm{cool}$, but in our case, this is not what is happening.  When weak perturbations are swept up in the wakes and carried to larger radii, their free fall times (t$_\mathrm{ff}$) increase.   The downward trajectory of  $t_\mathrm{cool} / t_\mathrm{ff}$ is due to the denominators becoming larger.  The perturbations are not cooling and therefore, they do not condense.  When the gas detaches from the wake, it mixes with the ambient medium and reverts to the median line.  We observe this as $t_\mathrm{cool} / t_\mathrm{ff}$ bouncing.  This demonstrates that even looking only at the local $ t_\mathrm{cool} / t_\mathrm{ff} $ of gas perturbations without tracking its history can result in an erroneous conclusion.

Finally, we note that while the high resolution of the Romulus simulations makes it possible to \emph{begin} to see the 
effects of the perturbations in the CGM, what we observe is very likely only the tip of the iceberg. Much more improvement in resolution is needed to fully study the phenomena and the impact of complex interplay between processes like turbulence compression and turbulent diffusion \citep[see, for example,][and references therein]{Rennehan2019,Rennehan2021}; depositional growth via mass flow onto the cloud due to radiative cooling and mixing-induced cooling; and growth via coalescence \citep[see][]{FaucherOh2023CGMphys}.

\subsubsection{Outward flow} \label{outward flow}
Lastly, we consider the fourth sub-component of the \emph{pre-existing CGM}, shown in \arf{lime-green} in Figs. \ref{tcooltff3D-RG1-p1} $\&$ \ref{tcooltff3D-RG1-p2}.   This is gas that has not been part of the BGG's ISM in the past 5 Gyrs and at $z_\mathrm{a}$ it is moving radially outward.   This outward flowing gas is treated separately from the \emph{central galaxy wind} because by our definition, the latter is essentially outflowing ISM.

As seen in the last column of Fig. \ref{thermohistory}, this gas was initially moving inward via either the cooling flow or the rapidly cooling, infalling precipitation prior to $z_\mathrm{a}$. However, before it becomes part of the BGG's ISM,  it is pushed back outward into the CGM by AGN outflows.   This reversal appears as a change of slope at $\Delta t =0$ in the top right panel of Fig. \ref{thermohistory}. After $z_\mathrm{a}$, this component essentially behaves like the \emph{pre-existing CGM}.

The total amount of gas involved and its outward velocity varies from halo to halo.   It is generally more pronounced in lower mass halos but is also affected by the timing and the strength of the last AGN outburst.   


\begin{table}
\fontsize{9}{11}\selectfont
\centering
\begin{tabular}
{|l|p{1cm}|p{0.5cm}|p{1cm}|p{1.5cm}|p{1cm}}
 \hline
 \hline
 \\
 Halo ID & CC/NCC status & $z_\mathrm{a}$ & Cooling flow & Precipitation & SFR [M$_\mathrm{\odot}$/yr]  \\

 \hline
 \\
 
C  & CC & 0.7  & 73.5\% & 28.2\% & 142.8 \\
52024  & CC & 0.29 & 61.7\% & 17.1\% & 73.9 \\
G1  & CC & 0.25 & 30.1\% & 25.0\% & 21.3\\
49510  & NCC & 0.36 & 2.1\% & 3.1\% & 0.0\\
99966  & CC & 0.31 & 71.5\% & 57.8\% & 22.3\\
38182  & NCC & 0.44 & 0.03\% & 0.0\% & 0.0 \\
91655  & NCC & 0.26 & 1.2\% &3.1\% & 0.64\\
77876  & NCC & 0.25 & 14.1\% & 4.1\% & 8.6\\

 \hline
\end{tabular}
\
\caption{Cooling flow and precipitation are the two main channels through which gas in the \emph{pre-existing CGM} cools. The table shows what fraction of gas that cools via these processes contributes to the star formation of the central galaxy or accretes to the SMBHs. The last column is the SFR of the central galaxy in each halo. The NCC groups generally have much less contribution to SF from cooling gas and, also have, much smaller SFR compared to the CC groups.}
\label{table:accreted-gas}
\end{table}

\subsection{Cool core vs. non-cool core groups} \label{preexisting-CCNCC}

At the end of \S \ref{tuCGM} we discussed the properties of CC and NCC groups and indicated that the r > 0.1$\rm R_\mathrm{500}$ gas properties of these groups are very similar. This similarity is also evident in the evolutionary behavior of the ``condensation-susceptible'' CGM analyzed in this section. The evolution of the seven patterns discussed above is largely the same regardless of whether the group is CC or NCC. The main differences are in the mass fraction of \emph{pre-existing CGM} that is cooling via cooling flow and precipitation, and directly contributing to the fueling star formation and the central SMBH when reaching the halo center. These are listed in Table \ref{table:3Bpre-existingCGM} and \ref{table:accreted-gas}.

In {\sc Romulus} groups,  30-50\% of the ``condensation-susceptible'' \emph{pre-existing CGM} in NCC groups behaves like cooling flow. In CC groups, this fraction increases to 60-70\%. G1, however, is an exception to this rule.  A more detailed analysis shows that G1 experienced a merger just before the start of our two Gyr window. It is just settling into the CC state and this transition is affecting its properties.

The next most important subcomponent of the \emph{pre-existing CGM} in the NCC group is the outward flow, comprising $\geq$ 35\%. This subcomponent moves outward because of its interaction with winds and AGN outflow. One of the highest fractions of outward flow belongs to halo 91655, which is undergoing an AGN outburst at $z_\mathrm{a}$, as seen in last row of  Fig. \ref{projections01}. The gas mass fraction for this subcomponent is lower in CC groups ($\sim$ 20\%), likely because CC groups have lower AGN activity at $z_\mathrm{a}$ than NCC groups.

As for the strongly perturbed (condensing) component, CC groups have $\sim$ 10 times more (in mass fraction) compare to the NCC groups, with G1 again being an exception. We note, however, that the fraction of condensing gas is, in general, small compared to all other components; this is expected since it is a phenomenon that requires ultra-high resolution to be resolved \citep{hummels2019impact,Das2021shatter,mandelker2021thermal}. In these {\sc Romulus} simulations, we are only just beginning to see the ``tip of the iceberg'' of this subcomponent.

All in all, an average of 65\% of ``condensation-susceptible'' gas is flowing inwards in CC systems (again G1 is an exception), while in NCC it is $\sim$ 40\%. Cooling flow and precipitation are the two main channels through which gas in the \emph{pre-existing CGM} cools. Table \ref{table:accreted-gas} shows what fraction of gas that cools via these two cooling channels contributes to fueling star formation or SMBH when it reaches the halo center (we have not distinguished between fueling star formation or SMBH in this table).  
A much higher fraction of the gas that cools via the cooling flow and precipitation in CC groups ($\sim$65\% and $\sim$30\%) contributes to star formation than NCC ($\sim$3-4\% and $\sim$3\%). Correspondingly, the SFR in CC systems is much higher. In two NCC systems, halos 38182 and 49510, there is no detectable star formation activity. 

However, not all of the  cooling gas reaches the BGG. In cosmological simulations, AGN outflows and stellar winds play an essential role in preventing the overcooling of the gas and the formation of  higher than observed stellar mass. This is readily apparent in Figure 9 of \citet{tremmel2019introducing}. The  star formation history in galaxy groups closely follow the ups and downs in SMBH activity.   Every increase in the SMBH activity is correlated with a decrease in the SFR, and conversely, when SMBH activity is quiescence, the SFR increases.

\section{Summary and Discussion}\label{discussion}

\subsection{Summary}\label{summary}

We begin our discussion by summarizing our main findings. The aim of our study is to investigate the origin and the evolution of the CGM around massive central galaxies in group-scale halos.  In the present study, we define the CGM as all gas in the radial zone $0.1 \leq \rm R/R_\mathrm{500} \leq 1$,  and focus our attention on a 5 Gyrs time segment during which the region under consideration is not disturbed by massive satellites.  A summary of our main findings is as follows:

\begin{enumerate}[leftmargin=0.4cm,label=(\alph*)]
\item{

The CGM of massive galaxies in the {\sc Romulus} simulations is far from uniform.   It is threaded by filaments of inflowing cooling gas and gaseous tails from satellites.  It also has localized patches of rapidly cooling gas, regions of  shocked gas, conical bipolar outflows linked to active AGN outbursts, and hot, low-density cavities inflated by recent AGN outflows.   Not surprisingly, the gas at any given radius consists of a spectrum of coexisting gas phases whose cooling times and $t_\mathrm{cool} / t_\mathrm{ff}$ ratios typically vary  by $\sim 2$ orders of magnitude.

}

\item{

We investigate the origin of the CGM in the {\sc Romulus} groups halos and identify four key ``sources’’: \emph{central galaxy wind}, \emph{subhalo internal}, \emph{subhalo stripped}, and \emph{pre-existing CGM}. The fraction of gas from each of these categories varies from halo to halo. However, in all the halos, the first two make up only a minor fraction of the CGM while the largest contribution comes from the last category.  In halos with quiet merger histories over the time period under consideration, $>80\%$ of the gas is associated with \emph{pre-existing CGM} and $\lsim 10\%$ is associated with \emph{subhalo stripped}.  In halos with a high rate of minor mergers or major mergers before the start of the analysis window, the \emph{subhalo stripped} fraction rises to $30-40\%$ while the \emph{pre-existing CGM} drops to $\sim 60\%$.

}

\item{

We find no significant difference in the CGM properties of the CC and NCC
groups on scales $\rm R/R_{500} > 0.1$.  For instance, their scaled gas temperature and entropy profiles are similar.  The  differences between the two are in the central region. CC groups have temperature profiles that drop towards the center.  They also have a greater amount of gas flowing/falling onto the BGGs and a higher central star formation rate.    We find that all {\sc Romulus} groups cycle through the CC and NCC states, with the former typically occurring between AGN outbursts.   
}

\item{ 

We find that CGM is also very dynamic.  Apart from undergoing repeated heating and cooling cycles, it is also subject to ongoing weak-to-strong density perturbations.  However, unlike idealized simulations where the main sources of the perturbations are shocks and turbulence caused by AGN outflows, in \emph{realistic} cosmological systems, the dominant sources of the perturbations are wakes, debris of tidal interactions, and stripped gas tails of orbiting satellites. We observe some of these perturbations forming cold condensations.
}

\item{

With respect to the condensing gas, one of our key findings is that the formation of condensations is \emph{not} restricted to regions in the halo where the median hot gas $t_\mathrm{cool} / t_\mathrm{ff}$ ratio falls below 10 or 20 or even 30.   At the same time, we also observe regions where the median $t_\mathrm{cool} / t_\mathrm{ff}$ ratio is below a threshold with very little gas condensing out. This suggests that the median value of $t_\mathrm{cool} / t_\mathrm{ff}$ of the X-ray emitting gas is not the main measure of the CGM's susceptibility to condensation. In a realistic cosmological environment, the amplitude of the density perturbations is just as, if not more, important.

}

\item{ We focus on local regions within the CGM with $t_\mathrm{cool} / t_\mathrm{ff} \leq 20$ and investigate their evolution.  In choosing this criterion, we are guided by \citet{choudhury2019multiphase} who suggest that local density perturbations with $t_\mathrm{cool} / t_\mathrm{ff} \leq 10$  in a stratified diffuse CGM are prone to condense. We have increased our bound to $t_\mathrm{cool} / t_\mathrm{ff} = 20$ because of factor $\sim$ 2 uncertainties in theoretical assessments \citep{choudhury2016cold}. We find two distinct evolutionary behaviours:

\begin{itemize}
\item{Case I: The density perturbations cool rapidly.  The temperatures of the associated gas decrease to $\sim 10^4$ K; the densities increase by a factor of $\gsim 30$; their $t_\mathrm{cool} / t_\mathrm{ff}$ ratios drop  to very low values; and they condense to form cool clouds.  In due course, most of these cool clouds fall inward towards the central galaxy.} 
\item{Case II: Neither the density, temperature, entropy nor $t_\mathrm{cool}$ change much.  However, 
their $t_\mathrm{cool} / t_\mathrm{ff}$ ratios drop for a time,  even dropping below our threshold, and then bounce back, eventually settling around the median value for the hot gas.  Examining these perturbations more closely, we find that these are weak perturbations that are dragged along in the wakes of orbiting satellites and swept to larger radii.  The reason their $t_\mathrm{cool} / t_\mathrm{ff}$ drops temporarily is not because these perturbations are cooling rapidly but because their free-fall timescale is increasing.  In due course, this gas detaches and mixes with the ambient CGM.   The latter is the reason for the bounce in $t_\mathrm{cool} / t_\mathrm{ff}$.   The evolution of these weak perturbations show that \emph{the local value $t_\mathrm{cool}$/t$_\mathrm{ff}$ falling below threshold \emph{does not}, by itself, unambiguously signal runaway cooling of a perturbation.}}
\end{itemize}
}

\item{
We find that in cosmological simulations, the perturbations in the CGM beyond the very central region (i.e.~ $\rm R/R_{500} > 0.1$) are primarily caused by infalling/orbiting substructure.

}

\item{

Finally, we find that the flow of the CGM onto the BGG occurs via three key channels:   Firstly, a fraction of the gas removed from the satellites cools and settles in the central galaxy.   Secondly,  regions subject to strong density perturbations cool and fall in.   And, thirdly, we identify a radiatively cooling component that flows subsonically into the central galaxy.  We find that the latter is not a spherically symmetric feature.  Rather, the cooling gas flows inward via filaments.   Most of these inflows have non-zero angular momentum and are responsible for the re-emergence of the gaseous disks in rejuvenating BGGs \citep[cf.~][]{jung2022massive,Lagos2022}.
}

\end{enumerate}

Lastly, it is important to note that  we are able to \emph{begin to} observe multiphase structure of CGM and especially, the effects of density perturbations on this gas reservoir, because the {\sc Romulus} suite of simulations are among the highest resolution cosmological simulations available.  However, we acknowledge that we are only scratching the surface.  To fully capture these varied phenomena and their impact on both the group central galaxies and the groups as a whole, considerable improvement in resolution is still needed. 

\subsection{Comparison with previous studies}\label{comparing}

 Recently, a number of studies  have investigated the CGM around central galaxies in halos of different masses
 \citep{van2019cosmological,suresh2019zooming, hafen2019origins,peeples2019figuring,nelson2020resolving,esmerian2021thermal,stern2021virialization}.  Most of these investigations have been carried out using either the FIRE-2 galaxy formation model \citep{hopkins2018fire}
layered onto the GIZMO hydrodynamics code \citep{hopkins2015Gizmo}; FOGGIE galaxy formation model \citep{peeples2019figuring} layered onto Enzo \citep{bryan2014enzo}; or TNG50 \citep{pillepich2019first} and Auriga \citep{grand2017auriga} layered onto the AREPO code \citep{springel2010Arepo}.  All except \citet{nelson2020resolving} have limited their efforts to the CGM of Milky Way mass galaxies (M$_\mathrm{vir} \sim 10^{12}$ M$_\mathrm{\odot}$) or lower.  Additionally, most of the studies focus mainly on the origin of CGM. Only \citet{suresh2019zooming}, \citet{nelson2020resolving} and \citet{esmerian2021thermal} investigate the dynamics of the CGM.  In our study, we use {\sc Romulus} galaxy formation model, which is different in detail from the FIRE-2 and the TNG models, to probe the origin and dynamics of the CGM around BGGs. {\sc Romulus} simulations are run using Tree+SPH code {\sc CHaNGa} \citep{menon2015adaptive}, which itself is different from the AREPO and GIZMO hydrodynamic codes. Comparing the results from various simulations is challenging due to the differences in the galaxy formation models used and the hydrodynamical codes leveraged to solve fluid equations, not to mention details like the mass of the halos under study and even the criteria used to tease out the details. Nonetheless, we attempt a brief comparison to the relevant results from previous studies.

We find that the CGM of BGGs in {\sc Romulus} simulations is multiphase and dynamic, characterized by a large range of densities, entropies, $t_\mathrm{cool}$ 
and $t_\mathrm{cool} / t_\mathrm{ff}$.
As discussed in \S \ref{fullCGM}, both
 \citet{esmerian2021thermal} and \citet{nelson2020resolving} find the same.
 The multiphase nature of the CGM in the  {\sc Romulus} C halo (the highest mass halo in our analysis) has been discussed previously by \citet{butsky2019ultraviolet}. They find gas with a range of temperatures (10$^4$ < T < 10$^6$ K) at all radii. Upon examining this gas further, \citet{butsky2019ultraviolet} find that some of it is highly enriched gas, which they interpret as evidence that it is gas from satellite galaxies.  We explicitly show that this gas is an important contributor to the multiphase structure of the CGM --- and can make up as much as 30--40\% of the cool gas --- in not only {\sc Romulus} C but also other group halos.

Examining the origin of the CGM in more detail, we find that the \emph{pre-existing CGM} makes up the largest proportion (> 60$\%$) 
of the CGM, followed by gas removed from subhalos.  
The other categories, gas inside subhalos, and gas from the \emph{central galaxy winds},
make up only a small fraction ($\lsim 4\%$) of the total. Both \citet{hafen2019origins} and \citet{nelson2020resolving} also explored the origin of the CGM.  A detailed comparison with \citet{nelson2020resolving} is not straightforward due to differences in our classification criteria. \citet{hafen2019origins} use similar criteria as us but they study the CGM around galaxies with masses M$_\ast \simeq 10^{6} - 10^{11}\;$M$_\mathrm{\odot}$ and find that the relative contribution of the different sources varies with galaxy mass.  However, their highest mass systems at low redshift appear to be similar to our groups: the \emph{pre-existing CGM} contributes the highest fraction (> 60\%) to the CGM, followed by  gas from satellite galaxies and then the \emph{central galaxy wind}\footnote{\citet{hafen2019origins} refer to these categories as  \emph{smooth IGM accretion}, \emph{satellite wind}, and \emph{wind}, respectively.}. This is consistent with our results.

 Our analysis shows that the CGM cools via two main channels: a filamentary isothermal cooling flow and strong perturbations that cool, condense and then rain onto the central galaxy.  Unlike idealized simulations, we find that the dominant source of strong perturbations in cosmological systems are satellite wakes and tails.   We were initially concerned whether this feature in the {\sc Romulus} simulations were real, as opposed to numerical artefacts.   However, the fact that \citet{nelson2020resolving} and \citet{esmerian2021thermal} find similar behaviour gives us some confidence. Additionally, we also examined the neighbourhood of the rapidly cooling gas and confirmed that they are spatially correlated with the satellite tails.  We do not have sufficient resolution to distinguish the extent to which the cooling is radiative, mixing-induced or both in near-equal measure.  Like us, \citet{esmerian2021thermal} find that over 70$\%$ of the cold phase in the CGM around their simulated Milky Way-like galaxies are also the products of non-linear density perturbations caused by cosmological accretion, feedback-driven winds, and the debris of tidal interactions from the central and satellite galaxies.   \citet{nelson2020resolving} also conclude the same, that the cold CGM is formed via large-density perturbations whose initial seeds are either fragments of tidal debris, ram-pressure stripped tails of satellites or non-linear perturbations in the CGM induced by satellite galaxies.  We note that in the present study, we have not considered local perturbations in, for example, gas removed from the subhalos, only the induced perturbations in the \emph{pre-existing CGM}.  The former too will be subject to condense to form clouds.

Turning to the \emph{central galaxy wind} component, our analysis
shows that it does not play an important role in contributing to cold gas. Some of the gas in \emph{central galaxy wind} cools, but most of it mixes with the CGM. This finding is different from that of \citet{suresh2019zooming} who note that $\sim$ 70\% of the cold CGM in their simulations has been processed by the central galaxy and forms due to the rapid cooling of wind material interacting with the hot halo. 
This difference could be due to any number of details, including differences in 
(i) the mass of the systems: \citet{suresh2019zooming} study the CGM of a M$_\mathrm{halo}$ $\sim$ 10$^\mathrm{12}$ at z $\sim$ 2. 
(ii) hydrodynamics solvers used to run the two simulations; 
(iii) the simulations' resolution: \citet{suresh2019zooming} run has enhanced resolution in CGM, achieving gas mass and spatial resolutions of 2200 $M_\odot$ and 95 pc, respectively. 
(iv) differences in the treatment of sub-grid physical processes, like mixing: moving mesh schemes have been shown to be more effective than SPH in capturing small-scale mixing and in tracking perturbations at mixing interfaces \citep{Rennehan2019,Rennehan2021}; 
(v) how the two galaxy formation models handle galactic winds or other related physics: 
\citet{suresh2019zooming} specifically discuss that their results are sensitive to details of galactic winds treatments and changes in this treatment can result in little to no wind particles cooling and condensing.

\subsection{Variations and Caveats}
\subsubsection{Exploring the impact of varying $t_\mathrm{cool} / t_\mathrm{ff}$ threshold}\label{threshold} 

In \S \ref{fullCGM}, we noted that various theoretical analyses suggest that when a local perturbation’s $t_\mathrm{cool} / t_\mathrm{ff}$ falls below $\sim 10$,  there is a high likelihood that non-linear damping mechanisms will not be able to prevent the non-linear growth of thermal instability.  This value is not a ``threshold’' in the strict sense because the competition between damping vs. growth also depends on various environmental properties, including the shape of the gravitational potential, the entropy profile, the efficacy of thermal conduction, and whether AGN feedback is on or off \citep{voit2021graphical, choudhury2016cold, binney2009high}.  For example, \citet{choudhury2016cold} argue that the value can be a factor of $\sim$ 2 higher.   For these reasons, we adopted a value of $(t_\mathrm{cool} / t_\mathrm{ff})_\mathrm{threshold} = 20$; however, we have also repeated the analysis for $(t_\mathrm{cool} / t_\mathrm{ff})_\mathrm{threshold} =10$ and $30$. Specifically, we have assessed the impact of varying the threshold on (i) the sources of the CGM and the ``condensation-susceptible'' CGM and (ii) the dynamics of the ``condensation-susceptible'' CGM by computing the corresponding mass fractions.

Simply put, we find that changing the threshold does not alter the source categories --- \emph{pre-existing CGM}, \emph{subhalo stripped}, \emph{subhalo internal}, and \emph{central galaxy wind}. 
However, we find that the mass fractions associated with these source categories do change. Specifically, the mass fractions of \emph{subhalo internal, subhalo stripped, and central galaxy wind} increase. More precisely, the mass in the \emph{subhalo internal} category does not change because of its very low $t_\mathrm{cool} / t_\mathrm{ff}$; the fraction increases because of the total mass of the gas with $t_\mathrm{cool} / t_\mathrm{ff}$ below threshold \emph{does} decrease as the latter is lowered from 30 to 10.  As for  \emph{subhalo stripped} and \emph{central galaxy wind}, the associated masses decrease with the threshold but not as quickly as the mass of the gas below our threshold.  Only the mass fraction of \emph{pre-existing CGM} decreases.  In spite of this,  \emph{pre-existing CGM} and \emph{subhalo stripped} remain the top two categories, and the ratio of these two mass fractions depends on whether the group under consideration is active or quiescent, as we had found previously.

As for the dynamics, we find that all seven behavioural patterns --- \emph{central galaxy wind}, \emph{subhalo internal}, \emph{subhalo stripped}, \emph{cooling flow}, \emph{strongly perturbed} gas, \emph{weakly perturbed} gas and \emph{outward flow} --- are present regardless of the value of $(t_\mathrm{cool} / t_\mathrm{ff})_\mathrm{threshold}$ but again, the gas mass fraction associated with each of these varies. 
We find that lowering the threshold from 30 to 10, leads to an increase in the mass fraction of the cooling flow and strongly perturbed CGM, while the mass fraction of weakly perturbed gas decreases.  All of these variations are straightforward to understand.  The  $t_\mathrm{cool} / t_\mathrm{ff}$ of the strongly perturbed gas drops to fairly low values and therefore, changing the threshold does not significantly alter the total mass in this component.  However, lowering the threshold does, as noted, decrease the total mass of the gas below threshold. This drives the mass fraction up.  The mass of cooling flow gas decreases as the threshold is lowered but not as fast as the total and therefore, the mass fraction also increases. On the other hand, $t_\mathrm{cool} / t_\mathrm{ff}$ of the weakly perturbed gas has a lower bound and consequently, associated mass with $(t_\mathrm{cool} / t_\mathrm{ff})_\mathrm{bounce} < (t_\mathrm{cool} / t_\mathrm{ff})_\mathrm{threshold}$  decreases as $(t_\mathrm{cool} / t_\mathrm{ff})_\mathrm{threshold}$ is lowered.

In addition to varying the $t_\mathrm{cool} / t_\mathrm{ff}$, we also examine the impact of defining ``condensation-susceptible'' CGM using the criterion $t_\mathrm{cool} < 1$ Gyr.   From Figs. \ref{tcool}, it is apparent that condensing gas is not limited to regions where the median $t_\mathrm{cool}$ of hot gas is below this threshold, and there are also regions where the median $t_\mathrm{cool}$ is below the threshold, yet only a small amount of gas is condensing out. This once again highlights the critical role of strong local perturbations in the formation of cold, condensed gas.

A more detailed analysis shows that even with the $t_\mathrm{cool}$ < 1 Gyr criterion, the sources and behavior patterns are  the same as before,  only details, like the corresponding mass fractions, change. The \emph{pre-existing CGM} and \emph{subhalo stripped} categories are still the top two categories and their respective mass fraction depends on whether a halo is active or quiescent, as discussed before. However, in one of our larger active halos, the \emph{subhalo internal} mass fraction is larger than that of the \emph{subhalo-stripped} mass fraction at the time of analysis.  This is a transient state.  As the orbiting substructures are stripped, this will change. 
As for the behavioral patterns, the fraction of cooling flow and strongly perturbed gas increases,  compared to the $(t_\mathrm{cool} / t_\mathrm{ff})_\mathrm{threshold} =10$ case, because their masses decreases at a slower rate than the total mass of the CGM with $t_\mathrm{cool}$ < 1 Gyr. On the other hand, the fraction of weakly perturbed gas decreases because the cooling time of  $~ 90\%$ of the perturbations is always longer than 1 Gyr and these are excluded by the $t_\mathrm{cool}$ < 1 Gyr criterion.

The most important outcome of these exercises is that our results regarding the origin of the CGM, the manner in which it evolves, and even our results concerning the cooling channels, are robust, independent of the specific choice of the threshold criterion adopted.

\subsubsection{Metal line cooling in {\sc Romulus} simulations}

Pushing to high resolution often requires making compromises.  In the case of {\sc Romulus}, a deliberate choice was made to treat only low-temperature ( $T \leq 10^4$ K) metal-line cooling.  This is explained in detail in  \citet{tremmel2019introducing, butsky2019ultraviolet} and \citet{jung2022massive}.   Here, we consider the potential implications for our study.

Metal lines collectively comprise the dominant radiative cooling channel for $T\sim 10^{5-7}\,\rm K$ gas. All things being equal, had full metal-line cooling been included in {\sc Romulus}, one would expect the CGM cooling time to shorten, which then would lead to (i) more gas with t$_\mathrm{cool}/\rm t_\mathrm{ff}$ ratio below the threshold, (ii) more massive cooling flow, and (iii) more gas condensing out.  In other words, the phenomena we have described in this paper ought be even more prominent.

However, this line of reasoning does not account for the fact that in real systems, stellar and  AGN feedback, act to offset cooling.  While this will undoubtedly alter exactly how much mass is associated with any one phenomenon, or how much gas is associated with cooling flows versus condensations, we assert that the general dynamics and the categories we have identified in this study are robust.

As for the details, although the precise fractions of cool and cold gas that forms in the CGM of massive galaxies are important for making observational predictions, these will have to await more realistic group simulations that are able to reproduce both the observed galaxy and the gas properties of groups. As discussed in detail in \S 2 of \citet{jung2022massive}, in current generation of cosmological  simulations, SMBH accretion and feedback sub-grid models are rather basic: They are tuned to offset cooling mainly to produce reasonably realistic distribution of galaxies.  In  simulations that include full metal-line cooling, a higher degree of cooling leads to more frequent and/or more energetic SMBH feedback episodes.  And while the CGM in these simulations is thermally balanced in a global, time-averaged sense, in detail the entropy profiles of the simulated groups typically have large, flat, high entropy cores \citep{oppenheimer2021simulating}.  Since observed groups have power-law entropy profiles \citep[][and references therein]{o2017complete}, we do not expect the detailed cooling/condensing profiles of these simulations to reflect that of real groups.
 
\subsubsection{Impact of Numerics: Resolution and Hydrodynamic Solvers}

We have already noted previously that investigations with enhanced CGM resolution show that as the resolution becomes increasingly finer, the CGM's warm/hot gas content decreases and its cool/cold gas content grows; this cool/cold gas fragments to progressively smaller sizes; and these clouds survive for a longer time
\citep{hummels2019impact,suresh2019zooming,van2019cosmological,peeples2019figuring}. At the same time, it is not clear how improving resolution will impact the modeling of, and the interplay between, processes like turbulent compression, turbulent mixing, and turbulent diffusion.  Some processes, like turbulent diffusion,  can limit fragmentation \citep[see, for example,][and references therein]{Rennehan2019,Rennehan2021} while others will exacerbate it.   These details will undoubtedly affect the precise amount of gas associated with different categories and dynamical behaviours.

Finally, we also consider the possibility that the  mass fractions may be sensitive to the numerical method used to solve the hydrodynamic equations, as suggested by the recent study by \citet{braspenning2022sensitivity}.   This study investigated the unfolding of cloud-wind interactions in simulations employing a variety of hydrodynamics solvers used in cosmological simulations. To isolate the influence of the hydrodynamic solvers, all simulations used the same initial conditions, were non-radiative, and did not include physical conduction.

In summary, they found a range of outcomes.  At one end of the spectrum, the trail of stripped gas from the primary cloud fragments into a swarm of dense, long-lived, cloudlets. At the other end of the spectrum, the gas forms a diffuse, swirling tail that rapidly mixes with the ambient medium.
This mixing, or lack thereof, is entirely numerical but we would expect that such differences will also impact the fractions of gas in the different components and states.   Moreover, we note that the inclusion of additional physical processes --- turbulent mixing, turbulent diffusion, physical conduction, radiative cooling, etc. --- will all also impact the outcome.  

Separately from the above issue, one can even question whether the condensing gas in our simulations is a numerical artefact.  Throughout this paper, we have been concerned about this.  However, we are assured that it is not, that at least it is not a product of {\sc CHaNGa}'s hydrodynamic solver.  After all, both \citet{nelson2020resolving} and \citet{esmerian2021thermal}, who use simulations that employ very different hydrodynamic solvers, also report condensing gas.

\section{Conclusions}\label{conclusion}

Our results highlight the complex and dynamic nature of CGM around massive galaxies in the {\sc Romulus} suite of high-resolution hydrodynamic cosmological simulations. The high resolution of {\sc Romulus} allows us to begin to see the evolving multiphase structure of CGM.  The key outcomes of our study are our results about the origin of the CGM and the manner in which the different components of the CGM evolve, including the two main cooling channels.  With respect to the latter, the CGMs flows onto the central BGG via a filamentary cooling flow as well as infalling cold gas.  We find that the latter are condensations that form from large density, ``condensation-susceptible'' perturbations induced mainly by orbiting satellites. In general, we find that these satellites trigger both strong and weak perturbations and the high resolution {\sc Romulus} simulations allow us to witness the evolution of both classes of perturbations. The $t_\mathrm{cool} / t_\mathrm{ff}$ ratios of both initially drop, and may even fall below threshold but in the case of the strong perturbations, this is due to a drop in $t_\mathrm{cool}$ and the perturbations rapidly cool.  In the case of weak perturbations, the drop is due to increasing $t_\mathrm{ff}$ and these perturbations do not condense. We also see the settling of gas stripped from orbiting satellites; on the average, however, this is sub-dominant.  

The patterns that we have described above are robust.  We also compute the fractions of mass associated with the different components.   Knowing these fractions are essential for making observational predictions.  However, as discussed,  there are too many uncertainties that have yet to be resolved, not to mention genuine large halo-to-halo variations that correlate with the merger history of the halos and with the cycle of AGN outbursts.

\section*{Acknowledgements}

We thank M.~Lehnert, B.~Oppenheimer, C.~Hayward, D.~Fielding, M.~Ruszkowski, G.M.~Voit, M.~Donahue, I.~Butsky, A.~Man, U.~Steinwandel, I.A.~Asensio, and B.~Keller for insightful discussions and suggestions.  VS, AB, and DR acknowledge support from the Natural Sciences and Engineering Research Council of Canada (NSERC) through its Discovery Grant program. DR also acknowledges support from NSERC through a Canada Graduate Scholarship (funding reference number: 534263). AB acknowledges support from the Infosys Foundation via an endowed Infosys Visiting Chair Professorship at the Indian Institute of Science.  AB, TQ, and MT were partially supported by NSF award AST-1514868. SLJ is supported by the Australian National University Research Scholarship.
MT is supported by an NSF Astronomy and Astrophysics Postdoctoral Fellowship under award AST-2001810. SIL is supported in part by the National Research Foundation of South Africa (NRF Grant Number: 146053). 
EOS acknowledges support from NASA through XMM-Newton award 80NSSC19K1056. And, PS acknowledges a Swarnajayanti Fellowship (DST/SJF/PSA-03/2016-17) and a National Supercomputing Mission grant from the Dept.~of Science and Technology, India.

The {\sc Romulus} simulation suite is part of the Blue Waters sustained-petascale computing project, which is supported by the National Science Foundation (via awards OCI-0725070, ACI-1238993, and OAC-1613674) and the state of Illinois. Blue Waters is a joint effort of the University of Illinois at Urbana-Champaign and its National Center for Supercomputing Applications. Resources supporting this work were also provided by the (a) NASA High-End Computing (HEC) Program through the NASA Advanced Supercomputing (NAS) Division at Ames Research Center; and (b) Extreme Science and Engineering Discovery Environment (XSEDE), supported by National Science Foundation grant number ACI-1548562. Analysis reported in this paper was enabled in part by WestGrid and Digital Research Alliance of Canada (alliancecan.ca). Our analysis was performed using the Python programming language (Python Software Foundation, https://www.python.org). The following packages were used throughout the analysis: numpy (\citealt{harris2020array}),  SciPy (\citealt{virtanen2020scipy}), and matplotlib (\citealt{hunter2007matplotlib}). This research also made use of the publicly available tools Pynbody (\citealt{pynbody}) and TANGOS (\citealt{pontzen2018tangos}).

Finally, VS and AB acknowledge the l{\fontencoding{T4}\selectfont
\M{e}}\'{k}$^{\rm w}${\fontencoding{T4}\selectfont\M{e}\m{n}\M{e}}n 
peoples on whose traditional territory the University of Victoria stands, and the Songhees, Equimalt and
\b{W}S\'{A}NE\'{C} peoples whose historical relationships with the land continue to this day. Similarly, SLJ acknowledges the Ngunnawal and Ngambri people as the traditional owners and ongoing custodians of the land on which the Research School of Astronomy \& Astrophysics is sited at Mt Stromlo.

\section*{Data Availability}

The data directly related to this article will be shared on reasonable request to the corresponding author. Galaxy database and particle data for {\sc Romulus} is available upon request from Michael Tremmel.



\bibliographystyle{mnras}
\bibliography{example} 





\label{lastpage}
\end{document}